\documentclass{aa}
\usepackage{graphicx}
\usepackage{natbib}
\bibpunct{(}{)}{;}{a}{}{,} %% AetA convetnion
\usepackage{color}
\begin{document}
   \title{The radial extinction profiles of late-type galaxies}
   \subtitle{}

   \author{Samuel Boissier
           \inst{1}
        \and
           Alessandro Boselli
	   \inst{2}
	\and 	
	   V\'eronique Buat
	   \inst{2}
	\and
           Jos\'e Donas
	   \inst{2}
	\and 
	   Bruno Milliard
	   \inst{2}
%\fnmsep\thanks{Just to show the usage  of the elements in the author field}
          }

   \offprints{S. Boissier}

   \institute{Carnegie Observatories, 813 Santa Barbara Street,
              Pasadena, 91101 California, USA \\
              \email{boissier@ociw.edu}
         \and
             Laboratoire d'Astrophysique de Marseille, Traverse du syphon,
             F-13376 Marseille cedex 12, France 
             \email{Alessandro.Boselli@oamp.fr;Veronique.Buat@oamp.fr;Jose.Donas@oamp.fr;Bruno.Milliard@oamp.fr}
             }

   \date{2004}

   \abstract{We have used UV (FOCA) and FIR (IRAS) images of six nearby late
   type galaxies to study the radial variation of the UV extinction
   (deduced from the FIR/UV ratio). We compare the UV extinction
   gradient with other extinction indicators (Balmer decrement) and
   search for a relation between the extinction, the metallicity and
   the gas surface density among our galaxies. We detect in our small
   sample a clear relation between extinction and metallicity. 
   These observed relations are used to calibrate an
   empirical recipe useful for extinction correction in the UV,
   visible and near-infrared images of late type galaxies.
   }

   \maketitle
%
%________________________________________________________________

\def\Ha{H$\alpha$}
\def\Hb{H$\beta$}
\def\Pa{P$\alpha$}

\section{Introduction}

The physical properties of late type galaxies, and in particular the
ones that are determined from the stellar emission can be studied only
after an accurate dust extinction correction.  It is thus crucial to
have a sound understanding of the dust extinction in late type
galaxies before studying nearby as well as high redshift
galaxies (especially since the latter are observed in the optical
in UV-rest frame, where the extinction is particularly severe).

The definition of an accurate model useful for correcting
the UV, optical and near infra-red data for the 
dust extinction in disc galaxies is particularly 
difficult because of several uncertainties:
i) little is known about the relative geometrical 
distribution of the dust and stars of various spectral types
and ages at small and large scales.
On small scales, young stars spend a finite time within 
the star forming regions before migrating out of them 
\citep[e.g.][]{charlot00,panuzzo03}. On large scales, 
the properties and distributions of HII regions 
in spiral discs vary with the morphological type \citep{kennicutt89}.
ii) the extinction law
(variation of the extinction with wavelength) presents large
variations for different lines of sight in the Galaxy, and different
shapes in the Galaxy and in the Magellanic Clouds, in particular in
the UV domain \citep[e.g.][]{savage79}. 
It is thus unclear
which extinction law should be applied.  
iii) Despite recent significant improvement in characterising the
extinction curve
\citep[e.g.][]{desert90}, the nature of the dust 
(composition, size distribution)
is still
poorly known.

There are however several observational ways to 
directly estimate
the dust extinction 
in well defined spectral ranges:
using the Balmer lines by comparing the observed to the expected
decrement \citep{lequeux81,mccall85,kennicutt92}, 
and in the UV from the FIR (Far Infra
Red) to UV ratio \citep[e.g.][]{buat96,calzetti00,panuzzo03,charlot00,boselli03},%.
or from 
the slope $\beta$ of the UV continuum
\citep[e.g.][]{calzetti94,meurer95,meurer99,bell02}.

The Balmer decrement gives a good measure
of the differential extinction between the lines outside of
the ionized region.
Its use however requires an accurate determination of the
underlying Balmer absorption. Ideally, one needs high quality
spectra and good resolution for this purpose.
In resolved galaxies, this method can only be applied 
to HII regions. Given their peculiar nature, 
the effects of age and geometrical distribution 
\citep[e.g.][]{charlot00,poggianti01,panuzzo03} make it
difficult to extrapolate the 
results to the galaxy as a whole or to other wavelengths.
\citet{calzetti94} proposed a relation to link the Balmer
decrement to the extinction in the UV and optical in 
starbursts, which 
was subsequently found not to be valid in normal galaxies
\citep{bell02,buat02}.

Various models \citep{witt00,panuzzo03} have shown that the FIR/UV ratio
is the most reliable measure of the UV attenuation, depending weakly
on the relative geometry of the stars and the dust,
the extinction law, or the nature of the underlying 
stellar population. This results from
the fact that the stellar population heating the dust (emitting in
the far infra-red) is the same population responsible for the UV emission.

\citet{buat96} used a radiative transfer model to derive
the extinction of 152 disk galaxies observed both in the UV and in the
infrared, using the FIR/UV ratio. They found relatively moderate 
extinctions (0.9 and 0.2
magnitude for early-type and late-type disk galaxies respectively).
It has also been used by
\citet{boselli03} in a large sample of 118 Virgo galaxies. They found an
average UV extinction of 1.28, 0.85, 0.68 for galaxies of type
Sa-Sbc, Sc-Scd, Sd-Im-BCD respectively. In association with 
a geometrical model, the FIR/UV ratio can be used to predict
the dust extinction at all wavelengths, as was done
in a simplistic way by \citet{boselli03}.

While the FIR/UV ratio has been used successfully in unresolved
galaxies, it has never been studied in spatially resolved
objects. While many profiles are available for nearby galaxies
(surface brightness at various wavelengths, gas, metallicity),
we do not know generally the radial distribution of extinction in normal
late-type spirals.

We propose in this paper to adopt the FIR/UV as an extinction
diagnostic for spatially resolved nearby galaxies.  This is
done by computing FIR and UV radial profiles, and deducing from them
reliable extinction profiles.  The FIR/UV ratio does not trace
extinction on small scales (because of geometrical and transfer
effects) but this problem is avoided by considering azimuthally
averaged profiles and working at relatively low resolution.  This
radial variation of the dust attenuation will be compared to the
frequently used Balmer decrement gradient.

Our next goal will be to give to the reader an empirical recipe for
correcting for extinction UV to near-infrared radial profiles (at least
in a statistical sense). This will be achieved by studying the dependence
of the dust extinction on other local properties that are likely to
affect the amount of dust: the metallicity and the gas surface density.

The data used are presented in Sect. 2: FOCA UV images of 6 nearby
galaxies with their IRAS FIR counter-part.
We also present
the gaseous profiles and HII region (abundances and
extinctions) data used in our investigation.
While our input data (UV and FIR fluxes, metallicity, gas surface 
density) tend to 
decrease with the radius, it is still unknown what is the radial variation
of the dust extinction (as traced by the FIR/UV ratio) and of the dust
to gas ratio.
In Sect. 3, we present the extinction profiles obtained in the UV,
and predicted at other wavelengths with the help of 
a simple model.
In Sect. 4, we compare the attenuation in the UV with the 
one derived from Hydrogen lines in HII regions.
In Sect. 5, we study the dependence of the extinction and the 
dust-to-gas ratio on the metallicity. We propose a simple prescription to
estimate the extinction profile in any galaxy with either a known abundance
gradient, or a blue absolute magnitude and scale-length.
Our most important results are summarised in Sect. 6.

\section{Data and methodology}

The first purpose of this work is to obtain reliable dust extinction
gradients from the FIR/UV ratio of resolved galaxies. This exercise is
however strongly limited by the lack of UV images of FIR IRAS resolved
galaxies,
and by the poor FIR spatial resolution of the IRAS images.  
Indeed the sample of galaxies large enough to be spatially resolved at
100 $\mu$m is quite small \citep{rice93}.

At the present time, only six galaxies satisfy these conditions.
In the UV, the images we will use were obtained by FOCA 
at 2000 \AA{} \citep[FOCA is described in][]{milliard91}. Their
characteristics are given in Table \ref{tableuv}.  IRAS images at 60 and 100
$\mu$m are available for 5 of them in the high resolution catalogue of
\citet{rice93}. For the last one (M100), we made an IRAS HIRES request
to the IPAC web
page\footnote{http:\slash \slash irsa.ipac.caltech.edu\slash 
IRASdocs\slash hires\_over.html}.
General properties and the integrated fluxes of the sample galaxies
are given in Table \ref{tablegen}.
\begin{table}
\caption{UV characteristics of the FOCA images.}
\label{tableuv}
\begin{tabular}{l r r r r r }
\hline
Galaxy & Resolution & Exposure    \\
	& arcsec & sec \\
(1)  &  (2)  &  (3)  \\
\hline
M33  & 20 & 9 x 150   \\
M51  & 12 & 4 x 300  \\
M81  & 12 & 11 x 150 \\
M100 & 12 & 4 x 200  \\
M101 & 12 & 4 x 300  \\
M106 & 20 & 4 x 150  \\
\hline
\end{tabular}
\end{table}
\begin{table*}
\caption{Properties of the target galaxies.}
\label{tablegen}
%{\small
\begin{tabular}{l r r r r r l r r r }
\hline
Galaxy & UV & H & 60 $\mu$m & 100 $\mu$m & $D$  & $Type$     & $D_{25}$   & $B_T$ & $12+log(O/H)$  \\
	& mag & mag & Jy & Jy           & Mpc &          & arcmin & mag & at center \\
(1)  &  (2)  &  (3)   &  (4)  &   (5)    & (6) & (7)     & (8)    &(9)  & (10) \\ 
\hline
M33    &  6.12 & 4.35 & 421.6 & 1102.3 & 0.70  & SAcd    & 70.8 & 6.27  & 9.08 \\
M51    &  8.88 & 5.65 & 106.1 & 266.8  & 8.40  & SAbc    & 11.2 & 8.96  & 9.38 \\
M81    &  8.97 & 4.09 & 42.7  & 161.9  & 3.63  & SAab    & 26.9 & 7.89  & 9.35 \\
M100   & 10.56 & 6.81 & 25.9  & 69.2   & 17.00 & SABbc   & 7.4  & 10.05 & 9.36 \\
M101   &  7.99 & 5.80 & 72.1  & 200.8  & 7.48  & SABcd   & 28.8 & 8.31  & 9.06 \\
M106   & 10.02 & 5.71 & 26.4  & 77.8   & 7.98  & SABbc   & 18.6 & 9.10  & 9.09 \\
\hline
\end{tabular} \\
%} 
{\tiny 
Column 1: Galaxy name.
Column 2: the UV magnitude from FOCA images integrated within the largest radius used
for this analysis.
The FOCA magnitude is defined as $m_{UV}$=-2.5 log($F$)-21.175,
 where $F$ is the flux in erg cm$^{-2}$ s$^{-1}$ A$^{-1}$).
Column 3: H band magnitude \citep{jarrett03}.
Column 4 and 5 : Flux densities (IRAS) at 60 and 100 $\mu$m in Jy \citep{rice93}.
Column 6 : Distance in Mpc. References: \citet{madore85}, 
 \citet{feldmeier97}, \citet{freedman01}, \citet{ferrarese00}.
 17 Mpc is adopted for M100, as a member of Virgo.
Column 7 : Morphological type (as found in NED\footnote{NASA/IPAC Extragalactic Database}).
Column 8 : Major axis diameter at the 25.0 mag arcsec$^{-2}$ isophote in B,
Column 9 : total B magnitude (columns 8 and 9 are taken from the RC3 catalogue).
Column 10: Central abundance (deduced from the abundance gradient extrapolated to
radius=0).
}
\end{table*}
These 6 galaxies do not constitute a complete sample in any sense, but
are the few nearby objects for which a resolved analysis of the ratio
FIR(IRAS)/UV(FOCA) is possible at the present time.

The determination of the radial variation of the FIR to UV flux
ratio is done after computing the independent FIR (60 and 100 $\mu$m
and UV radial profiles (see Sect. 2.4). The images of each galaxy
in the UV and at 100 $\mu$m are shown in Figs. 1 to 6 (respectively in the top-right and
bottom-right panels). The calculation of the extinction profiles
(the UV extinction profile is shown in the top-left panel of Figs. 1-6)
is reported in Sect. 3.

Once an extinction gradient has been obtained from the FIR/UV ratio, our
next goals are (a) to compare them with other extinction indicators
(b) to see whether the extinction within galaxies is related to other
entities like the gas density or the metallicity. If present, these
relationships would be extremely useful for a more accurate
computation of the effects of dust in galaxy evolution models in which
extinction is often simply estimated from gas densities and abundances
(e.g. Guiderdoni \& Rocca-Volmerange, 1987).

HII regions can provide us with additional information since their
spectroscopic observation can be used to have an independent dust
extinction estimate (through the Balmer decrement) and for the
determination of the metallicity gradient.  Data in HII regions have
thus been collected for each galaxy.  They are presented in the
middle column of Figs. 1-6 and discussed in Sect.s 2.1 (extinction)
and 2.2 (abundances).

We finally need gas profiles. They are presented
in the middle-left panel of figures 1-6 for each galaxies, and
commented in Sect. 2.3 (the UV extinction per hydrogen atom is
shown in the bottom-left panel of Figs 1-6).

\subsection{Extinction in HII regions}

We use 
the $A(H\alpha)$ extinctions as given by the Balmer
decrement in individual HII regions in each galaxy. 
The references for the 
data are given in Table \ref{tabledatac}

The H$\alpha$/H$\beta$ intrinsic flux line ratio is 
relatively constant, and has a value of 2.86 in case
B recombination \citep{osterbrock74}. The extinction in H$\alpha$
can be derived from the comparison of the observed
ratio to this intrinsic value, adopting a Galactic extinction
law and a dust screen geometry \citep{lequeux81}.
This observed ratio can also be affected by the underlying
stellar Balmer absorption, which can be as strong as the
emission in \Hb. For this reason, we decided to use
only the data in which
this underlying Balmer absorption had been taken into account, usually 
through
a standard correction of $\sim$ 2 \AA{}  
\citep[e.g.][]{mccall85}.

Whenever possible, the extinction has
been re-computed by comparing the observed H$\alpha$/H$\beta$ ratio to
the theoretical one of 2.86, and removing the Galactic component, in
an attempt to homogenize the data. 
The differences with the published values are
nevertheless small with respect to the scatter: 
the largest correction is
0.5 magnitude and the median 0.11.

An alternative dust extinction determination in HII regions
can by obtained by
the comparison of the radio continuum to the \Ha{} flux
\citep[e.g.][]{caplan86,lequeux81}. Since such data are not 
available for all our galaxies, we are unable to make this 
comparison at this time.

\begin{figure*} 
\begin{tabular}{c l }
& \includegraphics[clip,width=0.36\textwidth]{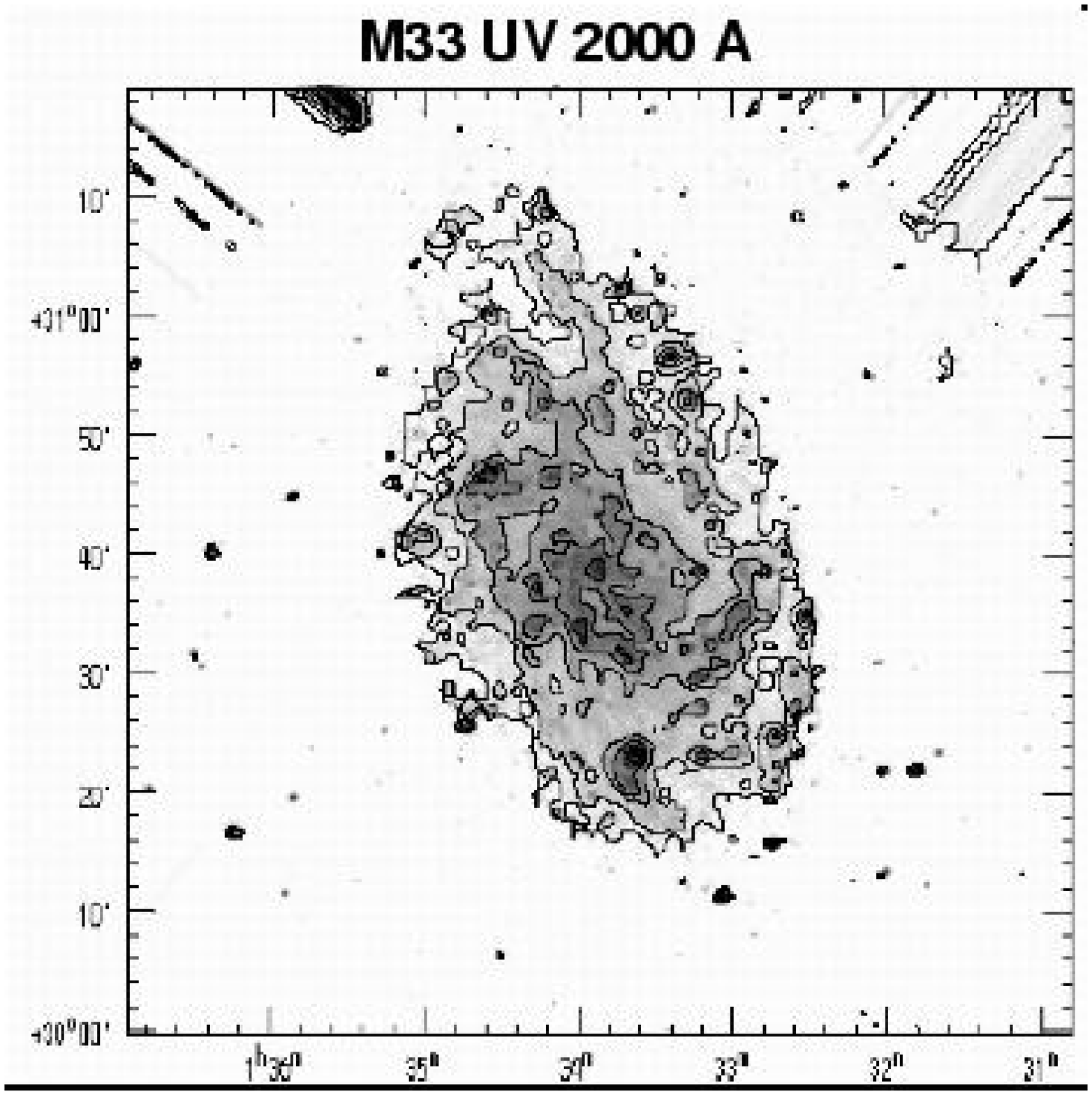}   \\
& \includegraphics[clip,width=0.36\textwidth]{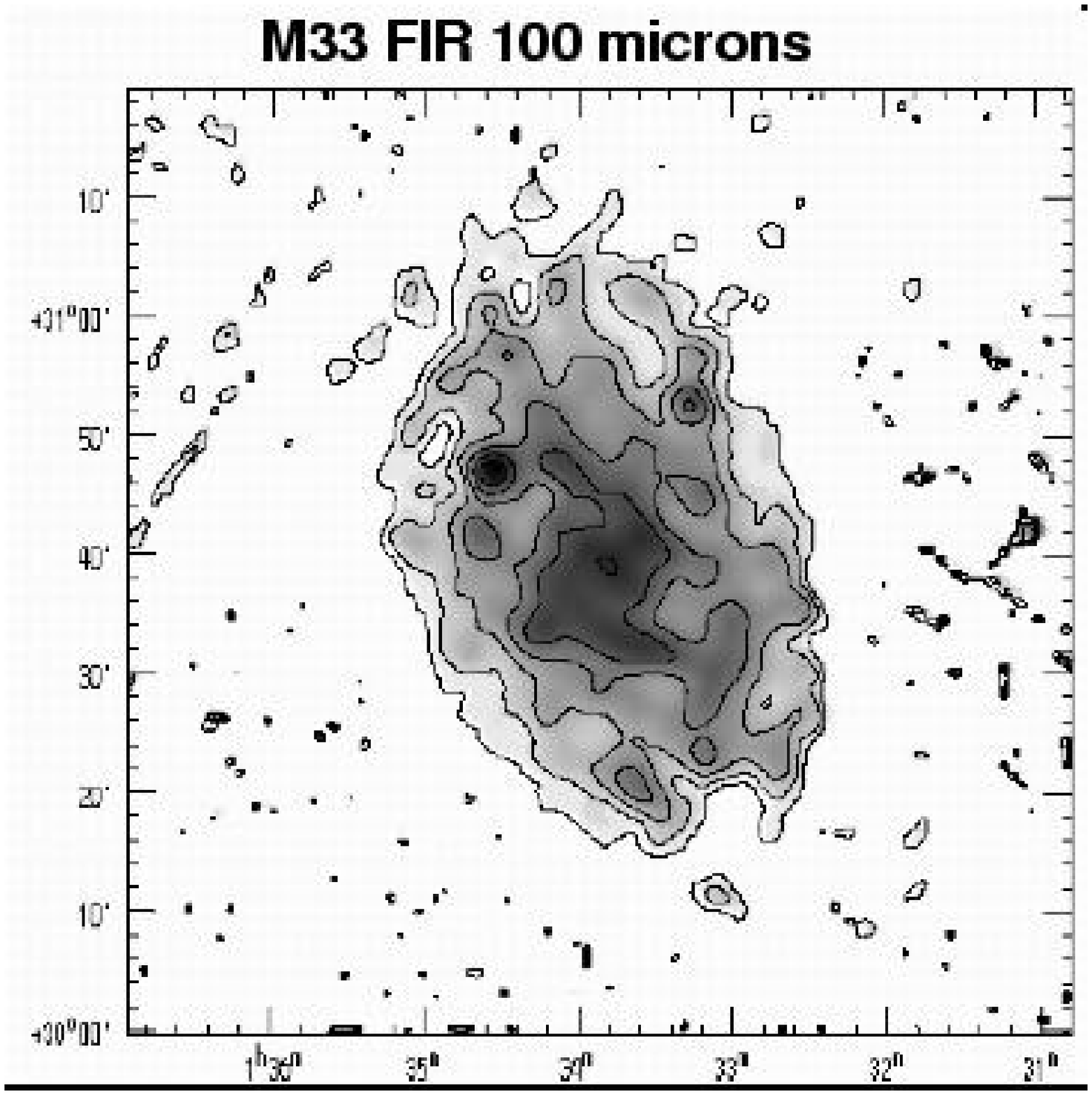} \vspace{-0.76\textwidth} \\
\includegraphics[width=0.58\textwidth]{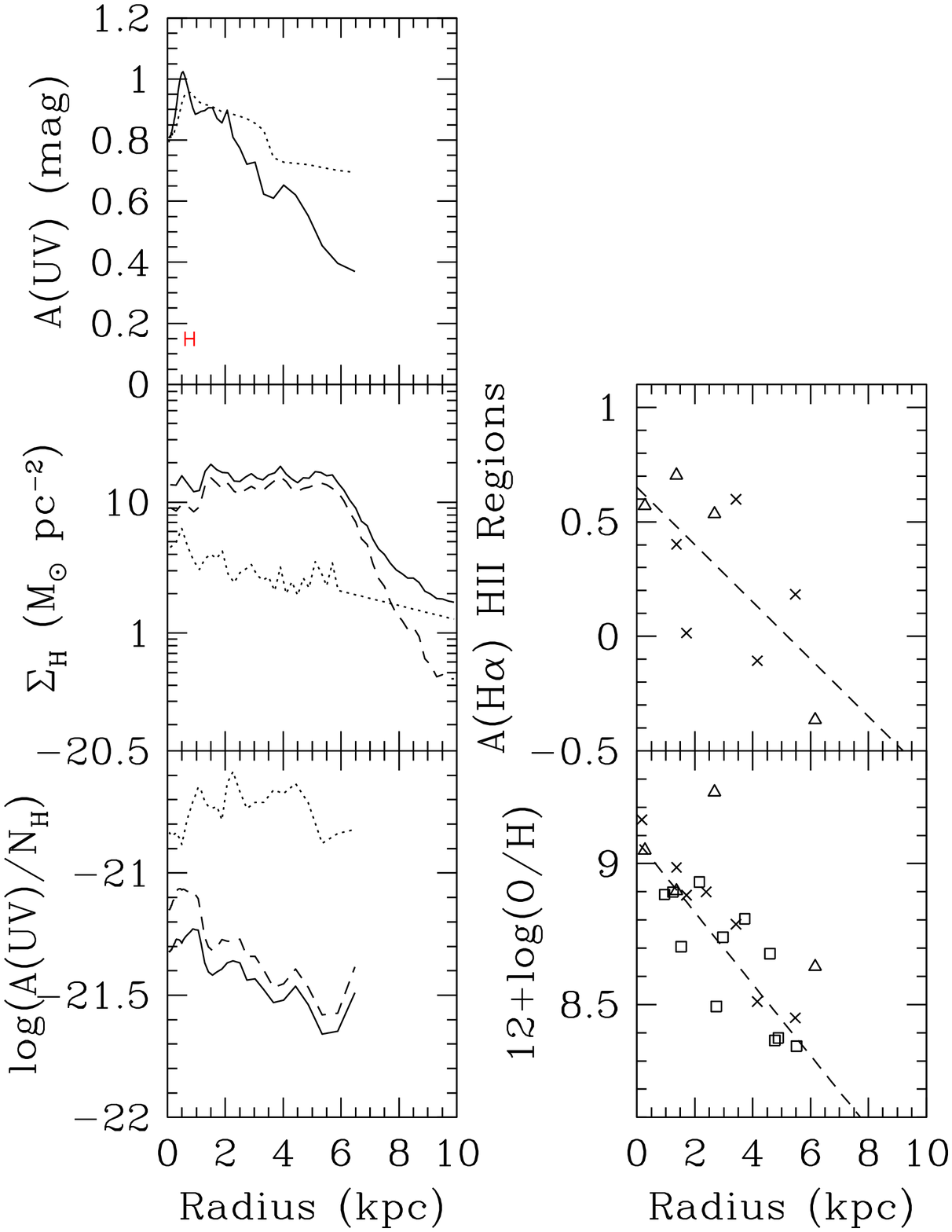} &
\end{tabular}
%\centering \includegraphics[width=0.5\textwidth]{individualM33.ps}
   \caption{Profiles in M33. 
\emph{\underline{Top row-left column}} : UV attenuation radial profile
   obtained from the FIR to UV ratio (solid). The dotted curve indicates the
   integrated extinction within the radius $R$. The error bar indicates
   the resolution (IRAS resolution at 100 $\mu$m). 
\emph{\underline{Middle row-left column}}: hydrogen density profiles 
(neutral: dashed, molecular: dotted, and total:solid). 
\emph{\underline{Bottom row-left column}}:
$A(UV)$ to hydrogen atom column density 
(neutral: dashed, molecular: dotted, and total:solid). 
\emph{\underline{Middle row-middle column}}: H$\alpha$ extinction in HII regions
(different symbols correspond to different data references
(Table \ref{tabledatac}), and fit 
(dashed line). 
\emph{\underline{Bottom row-middle column}}: oxygen
abundance in HII regions and fit (dashed line).
On the right, the UV (top) and IRAS 100 $\mu$m (bottom) 
surface brightness images are shown. 
The contours are separated by 1 mag arcsec$^{-2}$
in both images where $\mu$=-2.5 log($F$)+15 
($F$ in Jy arcsec$^{-2}$) for the 100 $\mu$m image
and $\mu_{UV}$=-2.5 log($F$)-21.175 
($F$ in erg cm$^{-2}$ s$^{-1}$ A$^{-1}$ arcsec$^{-2}$) for the 
2000 \AA{} image.
} \label{figM33}% 
\end{figure*}

\begin{figure*} 
\begin{tabular}{c l }
& \includegraphics[clip,width=0.36\textwidth]{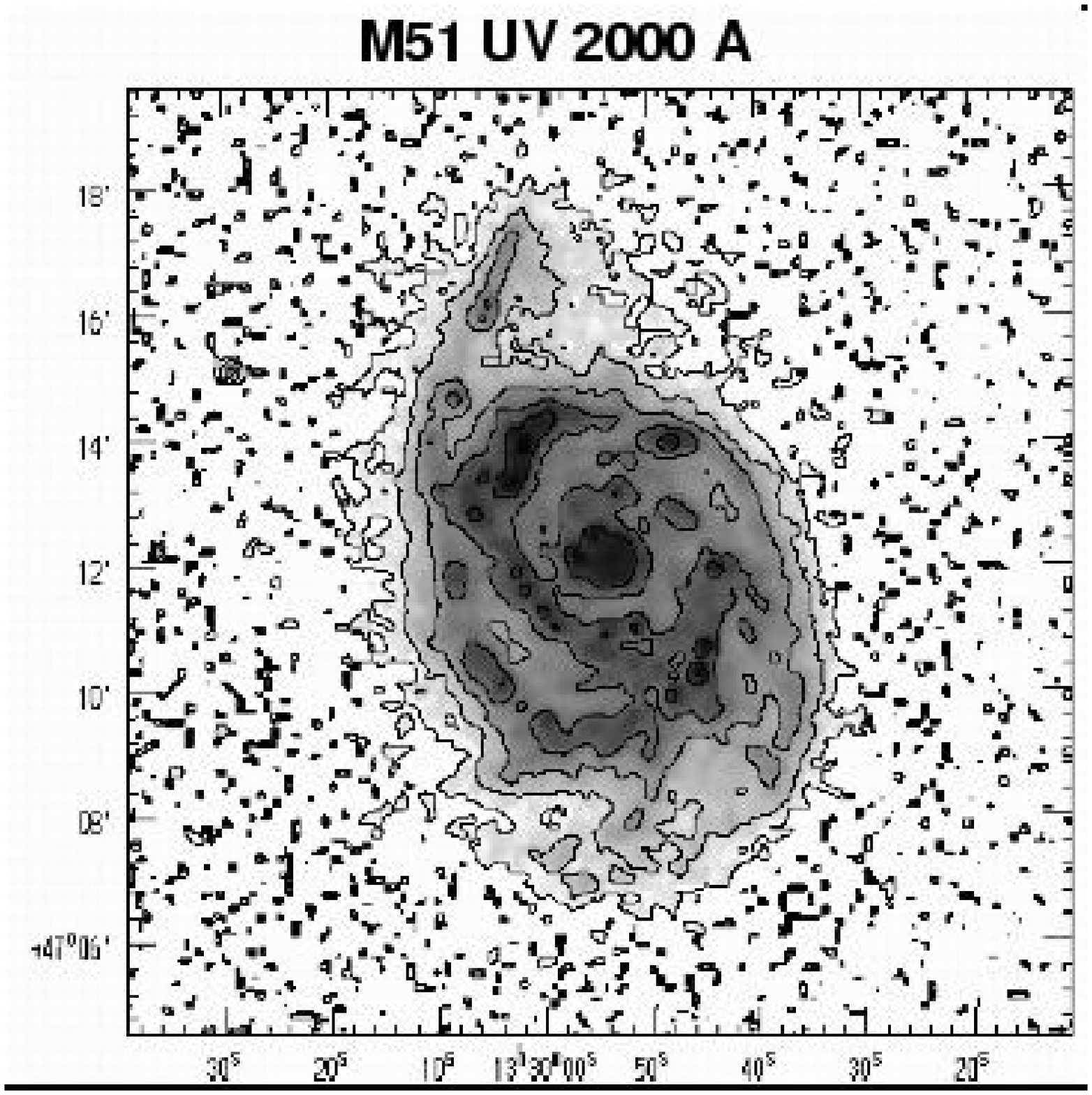}   \\
& \includegraphics[clip,width=0.36\textwidth]{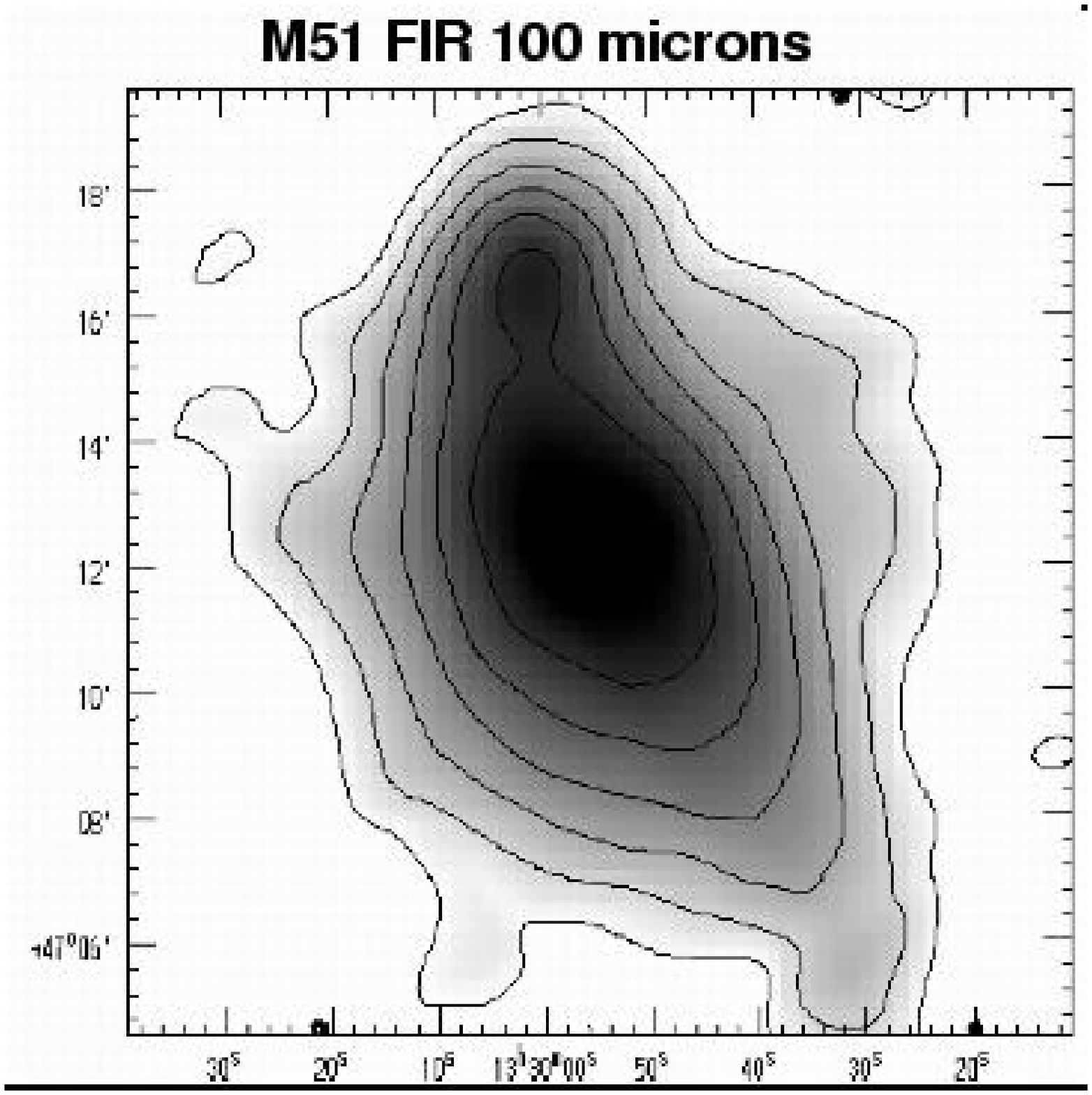} \vspace{-0.76\textwidth} \\
\includegraphics[width=0.58\textwidth]{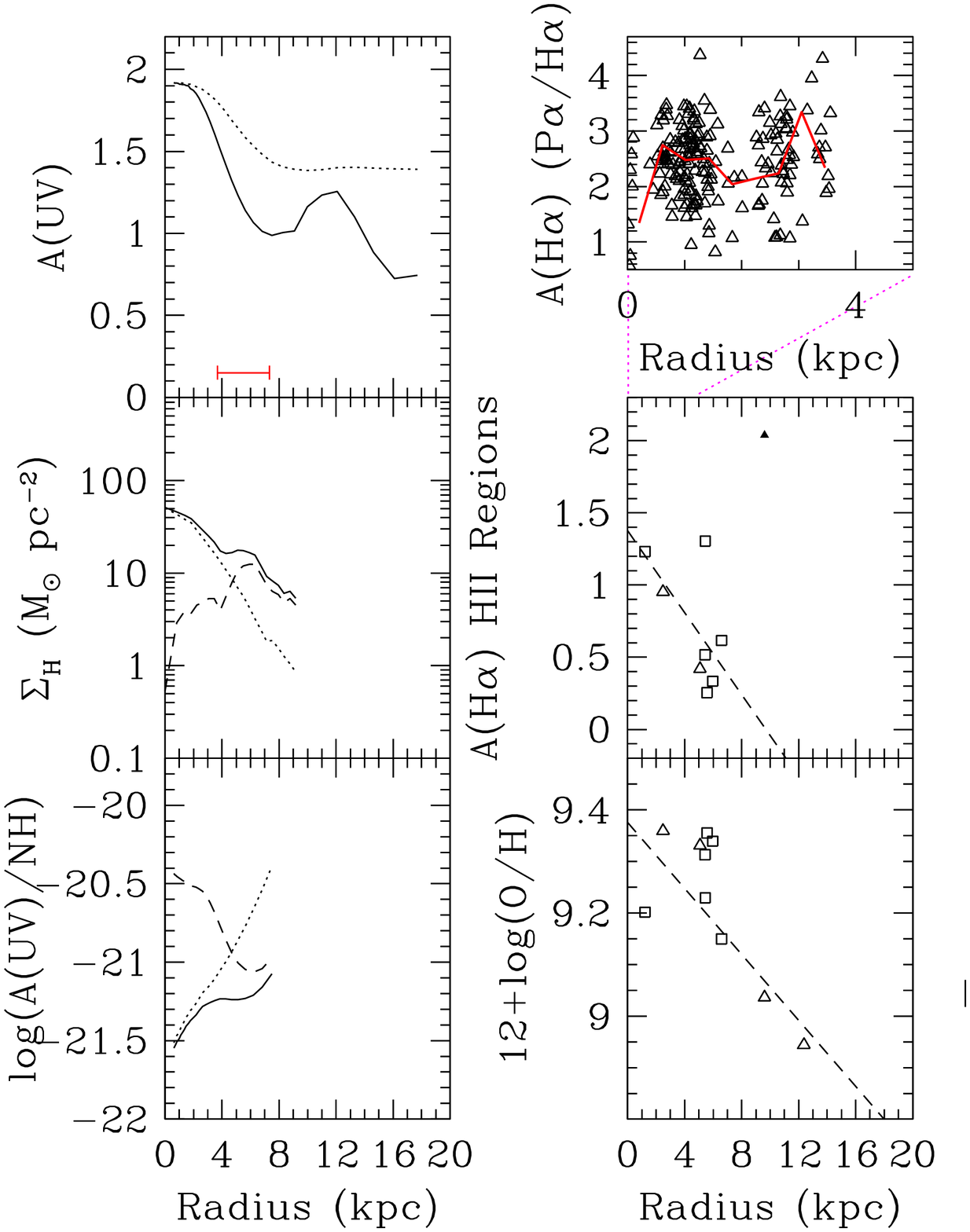} &
\end{tabular}
   \caption{Profiles in M51. 
Same caption as for Fig. \ref{figM33} except:
\emph{\underline{top row-middle column}}: H$\alpha$ extinction derived from Paschen-$\alpha$ to
H$\alpha$ ratio, individual HII regions from \citet{scoville01} (triangles).
The line is obtained by
averaging the \Pa{} and \Ha{} fluxes in 10 arcsecs radial bins, taking
into account only the pixels with signal to noise larger than 5 and 3 in
respectively the \Pa{} and \Ha{} images of \citet{scoville01}. 
} \label{figM51}%
\end{figure*}

\begin{figure*} 
\begin{tabular}{c l }
& \includegraphics[clip,width=0.36\textwidth]{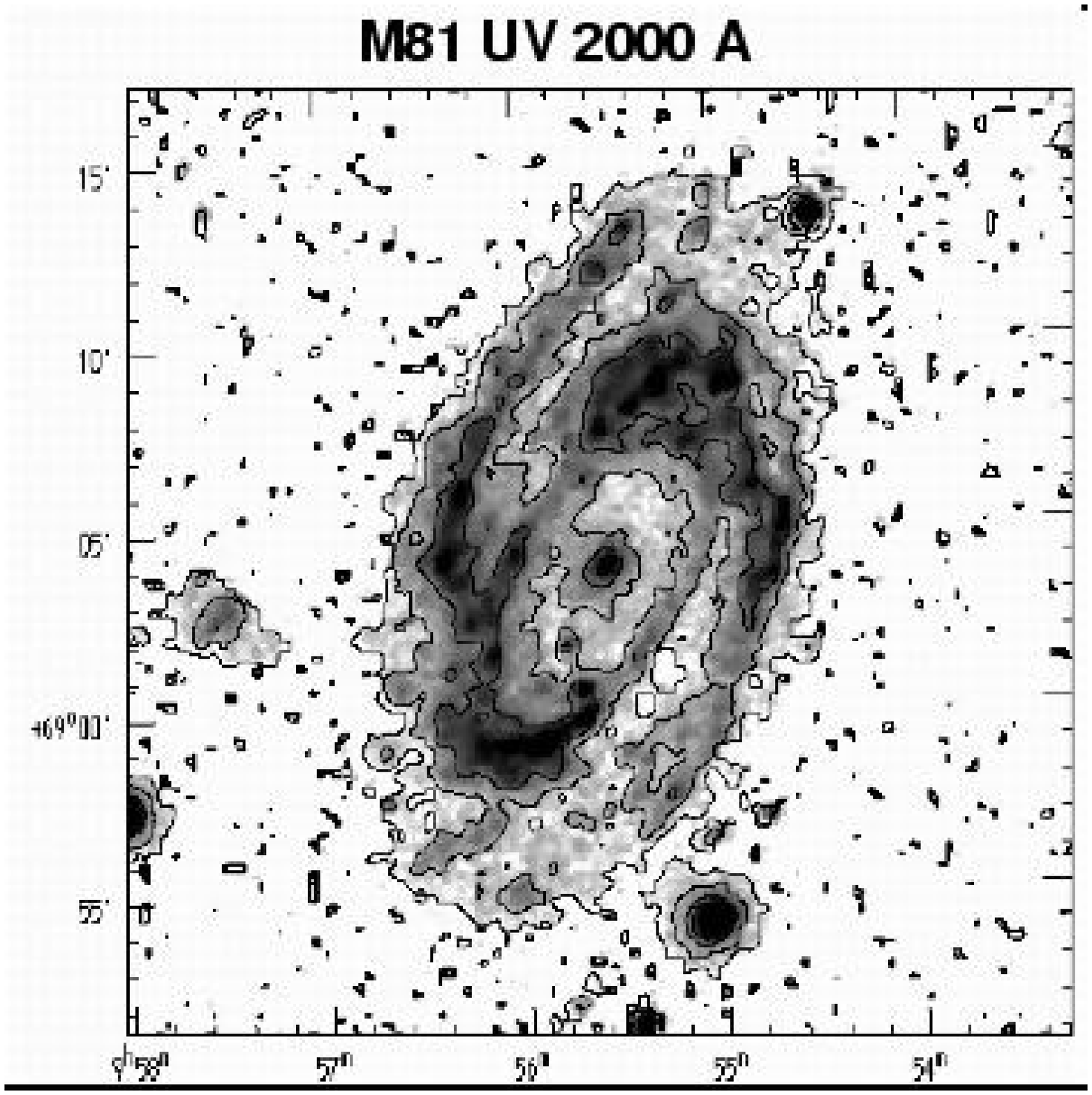}   \\
& \includegraphics[clip,width=0.36\textwidth]{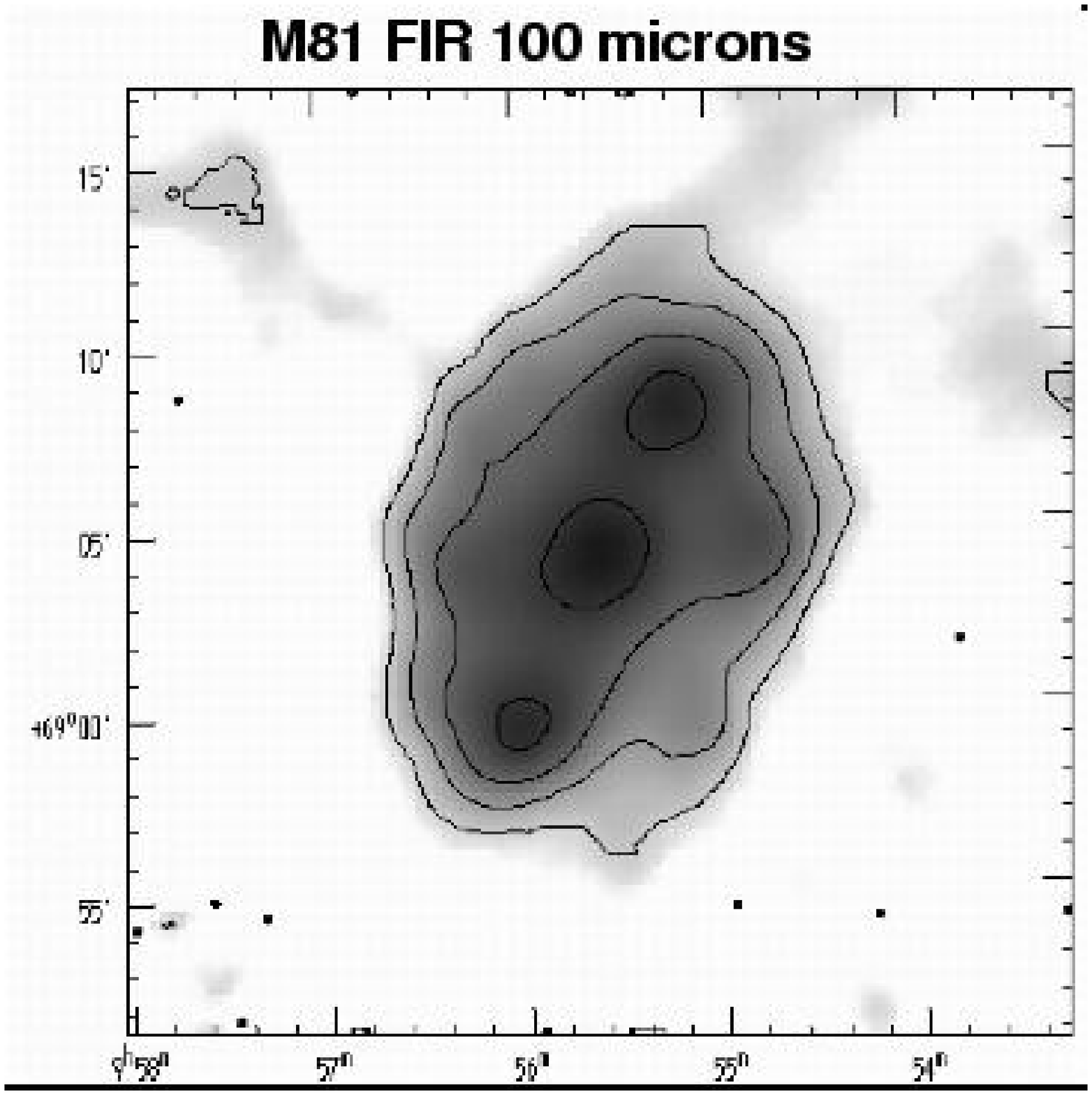} \vspace{-0.76\textwidth} \\
\includegraphics[width=0.58\textwidth]{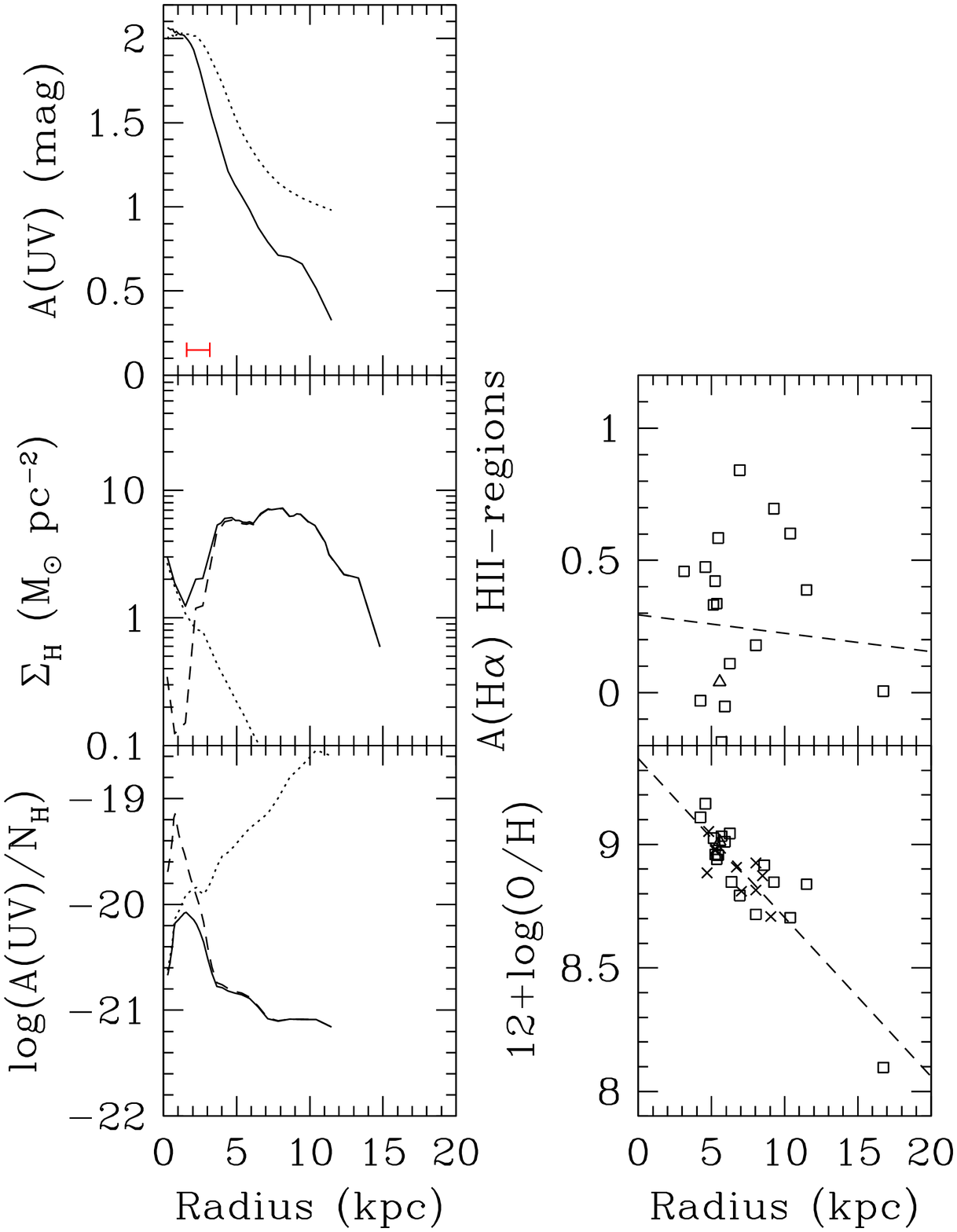} &
\end{tabular}
%\centering \includegraphics[width=0.5\textwidth]{individualM81.ps}
   \caption{Profiles in M81. 
Same caption as for Fig. \ref{figM33}.
} 
\label{figM81}%
\end{figure*}

\begin{figure*} 
\begin{tabular}{c l }
& \includegraphics[clip,width=0.36\textwidth]{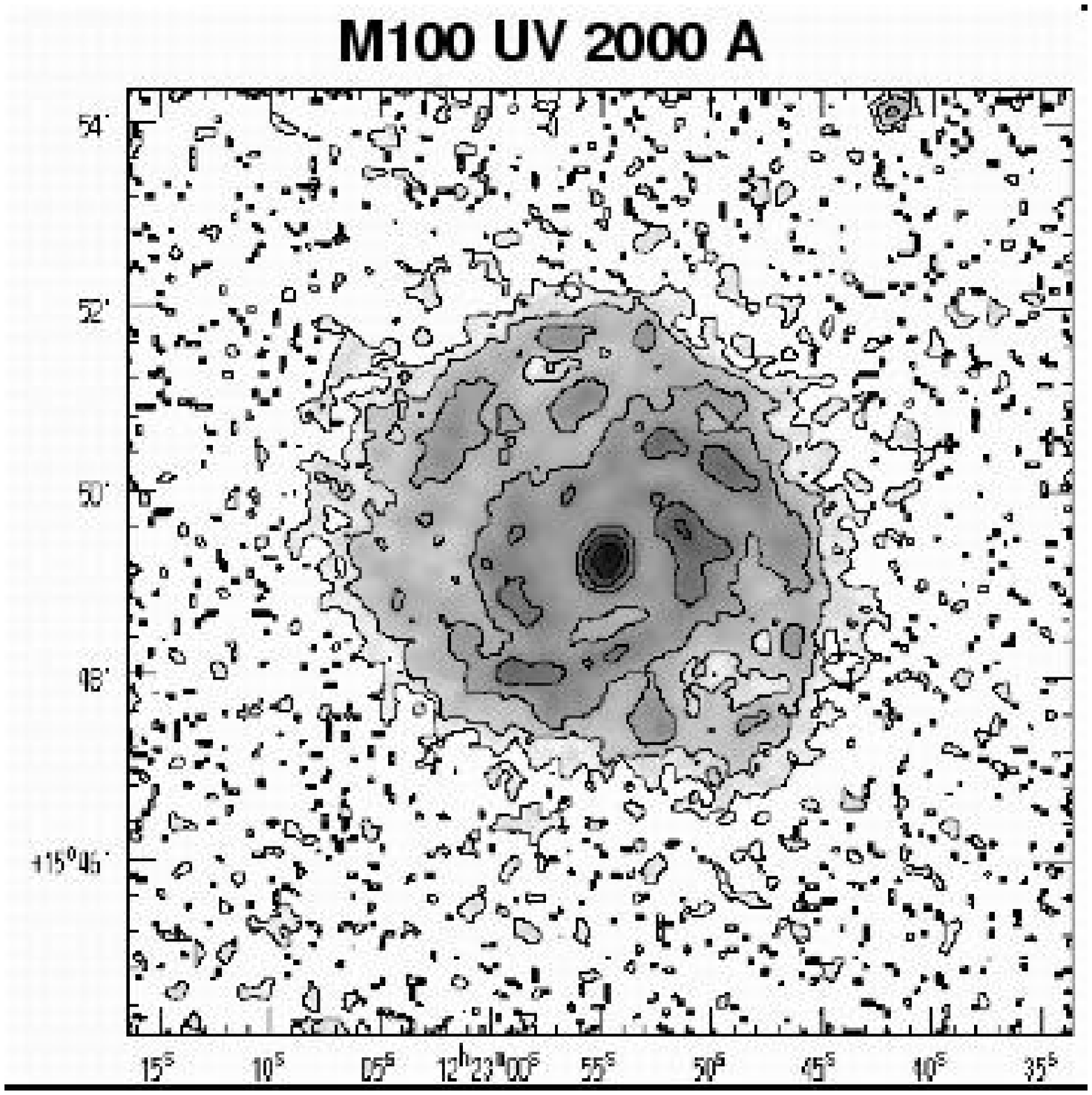}   \\
& \includegraphics[clip,width=0.36\textwidth]{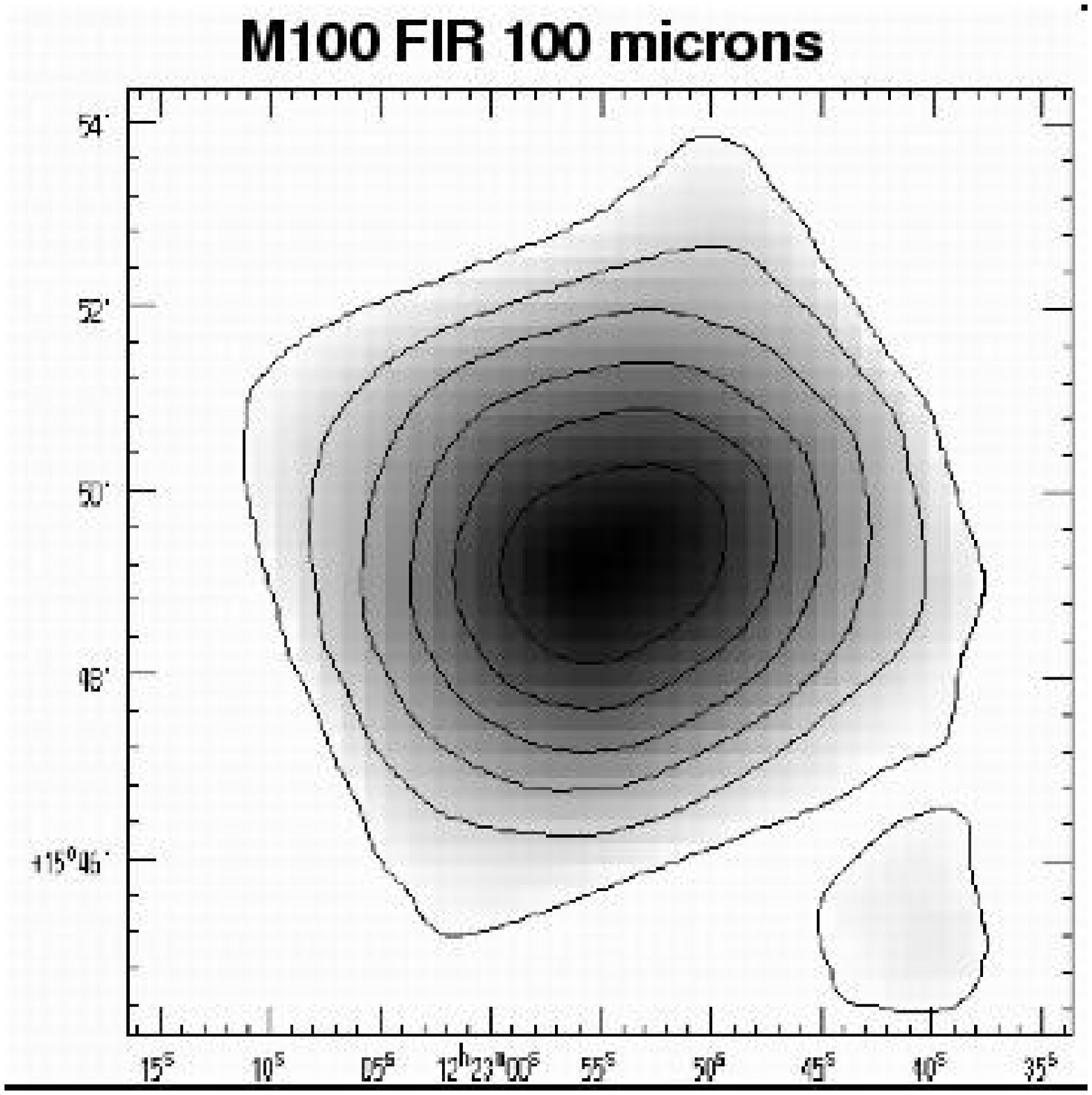} \vspace{-0.76\textwidth} \\
\includegraphics[width=0.58\textwidth]{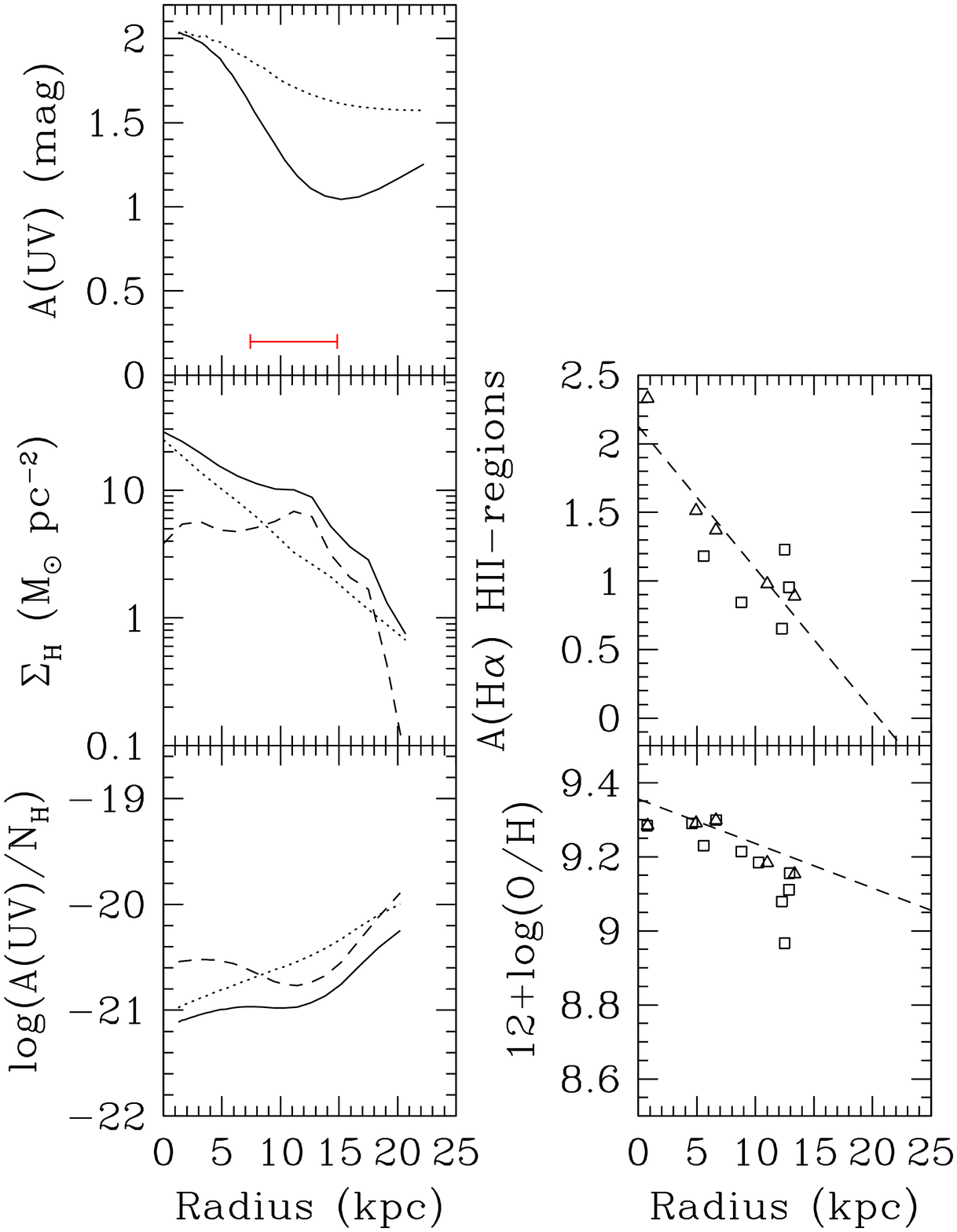} &
\end{tabular}
%\centering \includegraphics[width=0.5\textwidth]{individualM100.ps}
   \caption{Profiles in M100. 
Same caption as for Fig. \ref{figM33}.
} \label{figM100}%
\end{figure*}

\begin{figure*} 
\begin{tabular}{c l }
& \includegraphics[clip,width=0.36\textwidth]{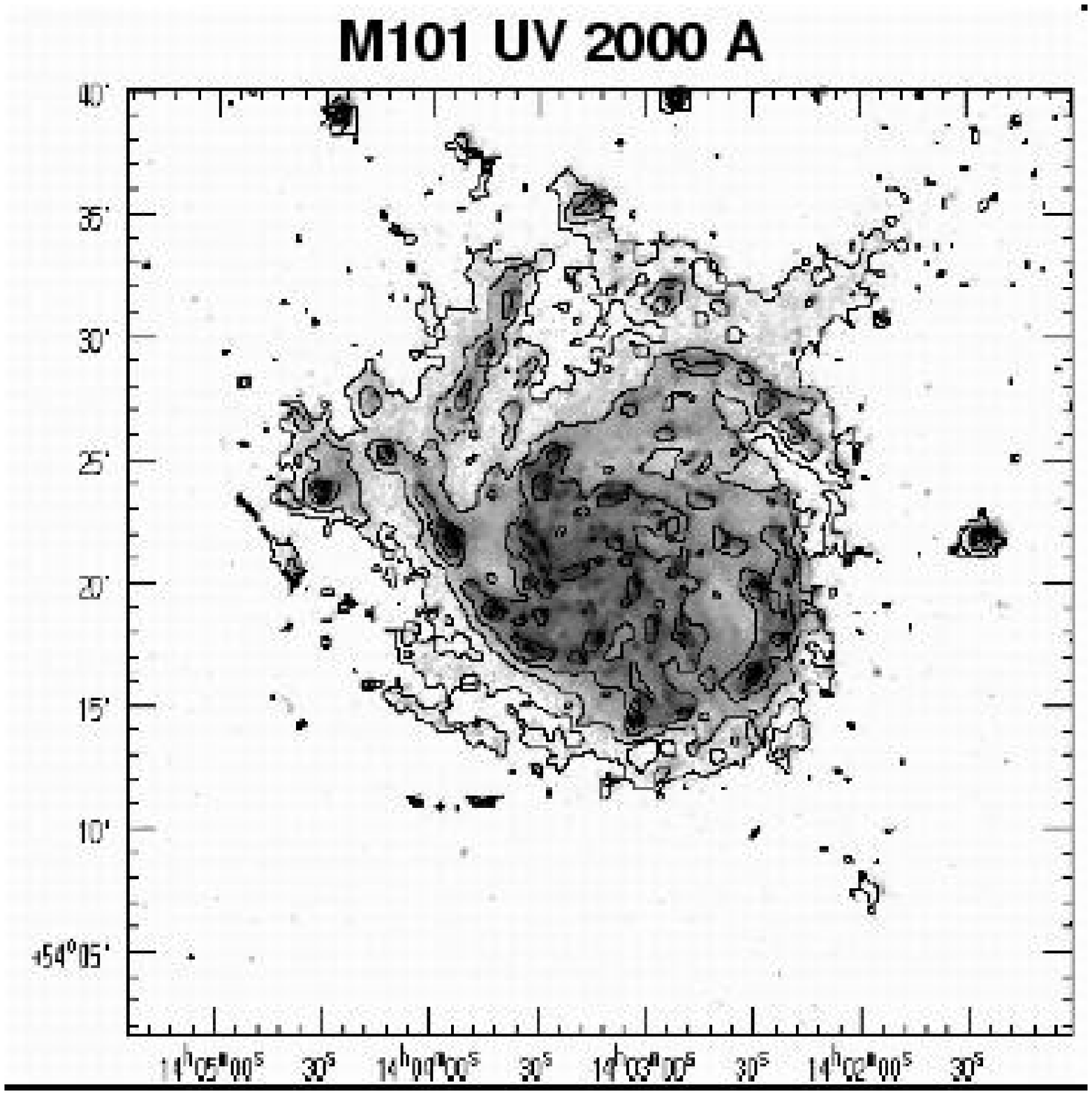}   \\
& \includegraphics[clip,width=0.36\textwidth]{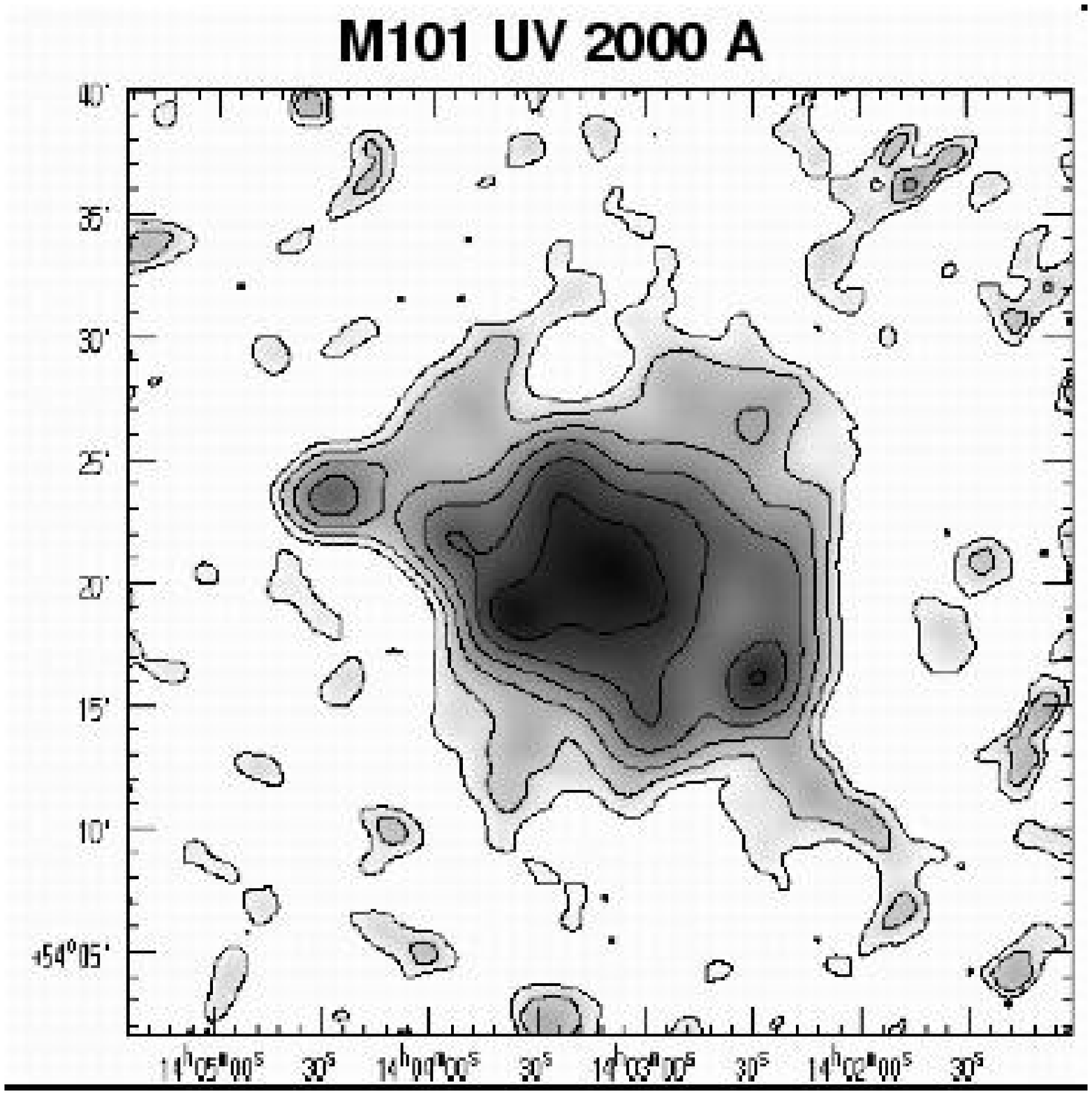} \vspace{-0.76\textwidth} \\
\includegraphics[width=0.58\textwidth]{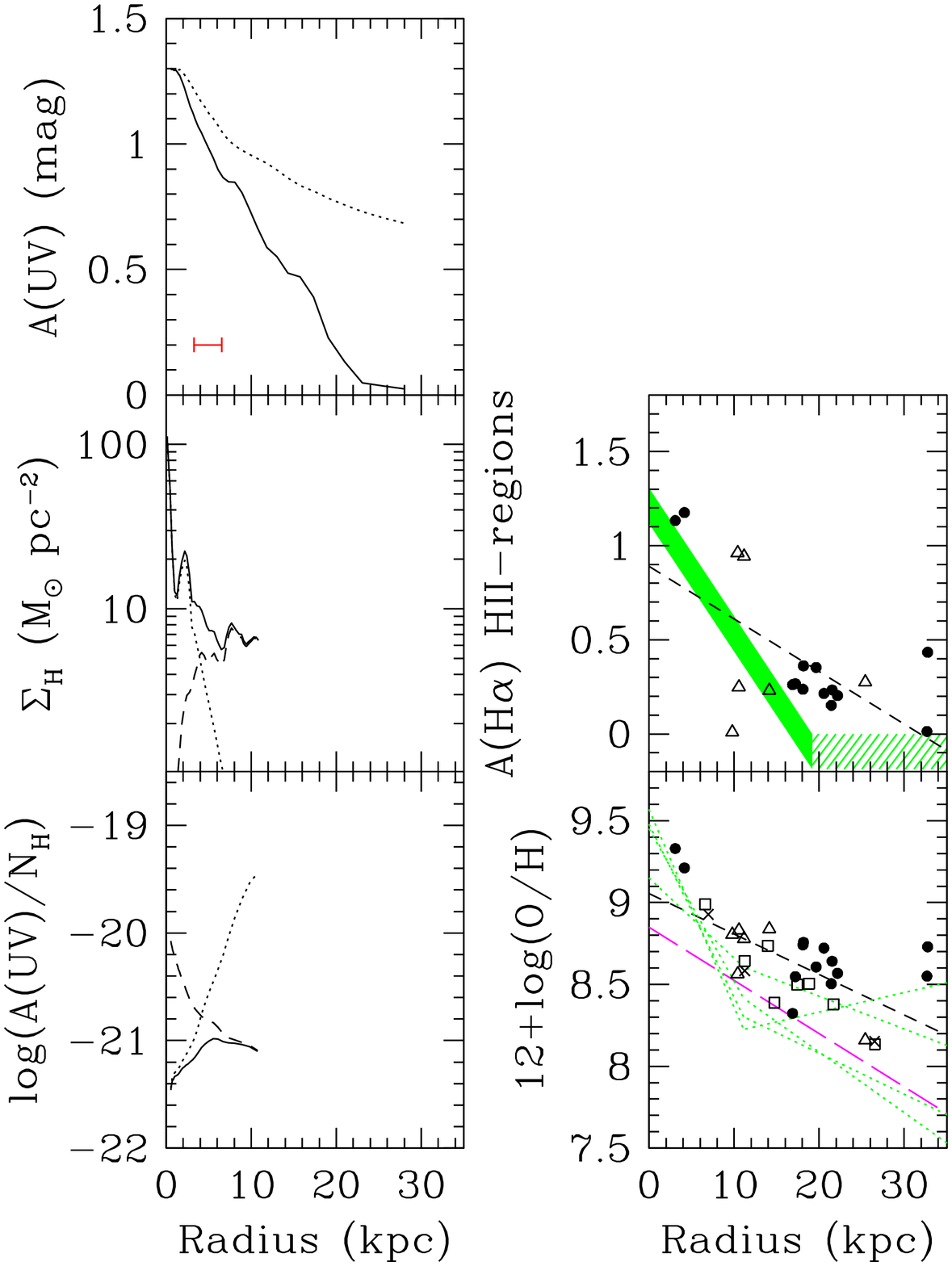} &
\end{tabular}
%\centering \includegraphics[width=0.5\textwidth]{individualM101.ps}
   \caption{Profiles in M101. 
Same caption as for Fig. \ref{figM33} except:
\emph{\underline{Middle row-middle column}}: The shaded area indicates roughly
the extinction gradient
derived by \citet{scowen92} from their narrow-band imaging study.
\emph{\underline{Bottom row-middle column}}: 
% oxygen abundance in HII regions and fit (dashed line). 
The dotted lines indicate the gradients of \citet{scowen92}
for the various values of the surface brightness threshold
they use with their narrow-band imaging.
The long-dashed line is the gradient of \citet{kennicutt03}
measured with high signal-to-noise data (and thus avoiding the use
of strong-line calibrations as in the other studies).
} \label{figM101}%
\end{figure*}

\begin{figure*}
\begin{tabular}{c l }
& \includegraphics[clip,width=0.36\textwidth]{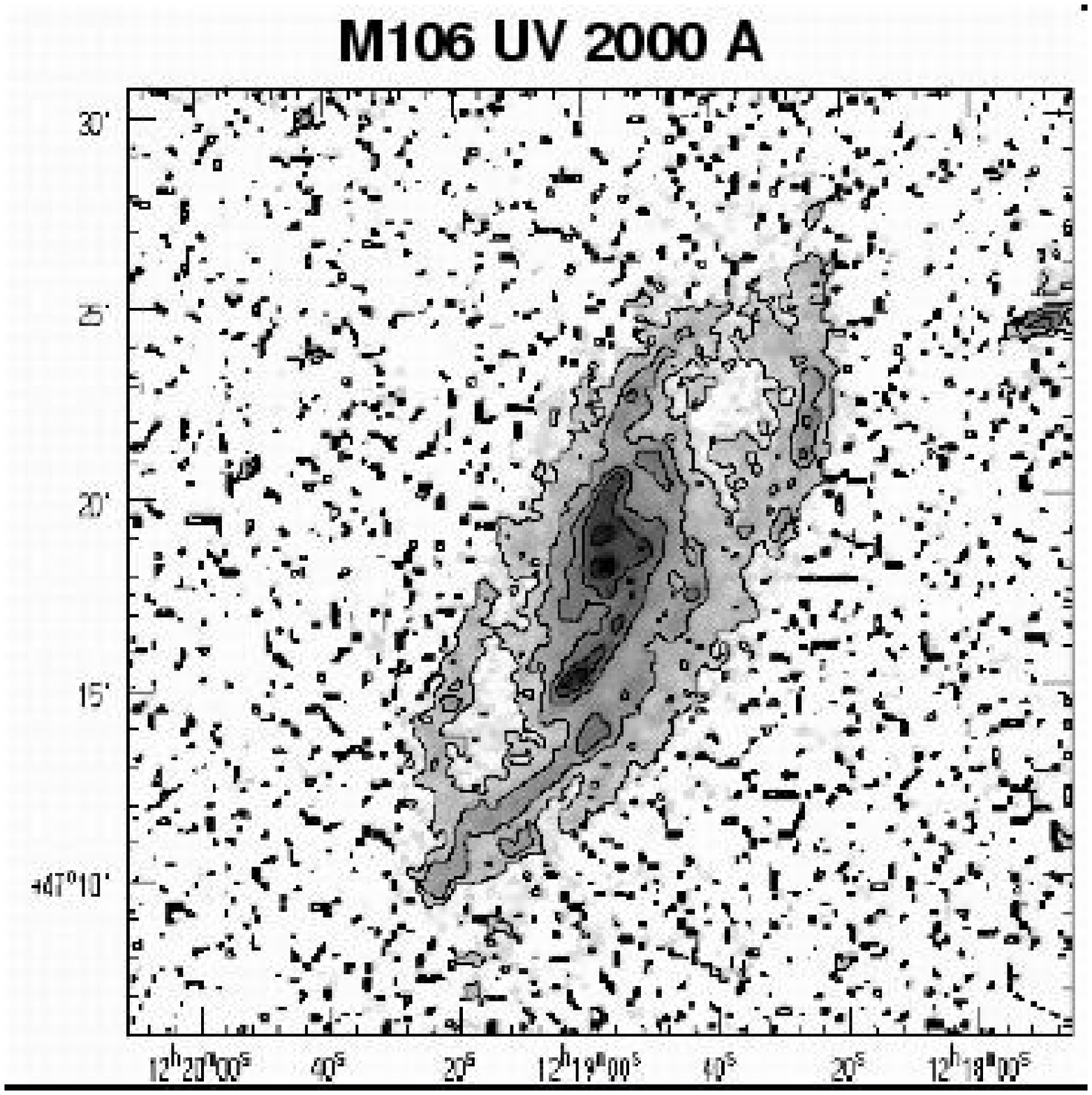}   \\
& \includegraphics[clip,width=0.36\textwidth]{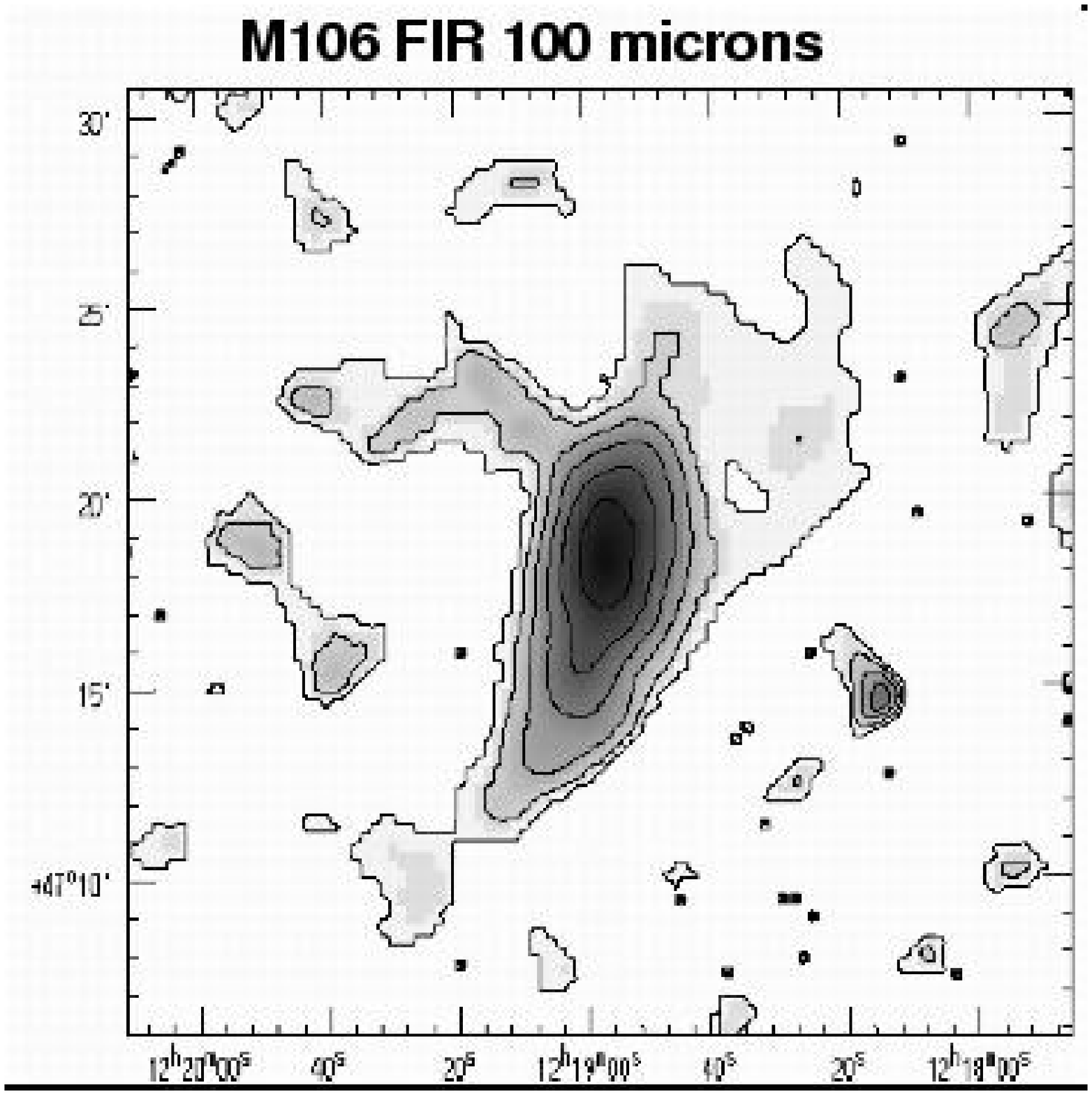} \vspace{-0.76\textwidth} \\
\includegraphics[width=0.58\textwidth]{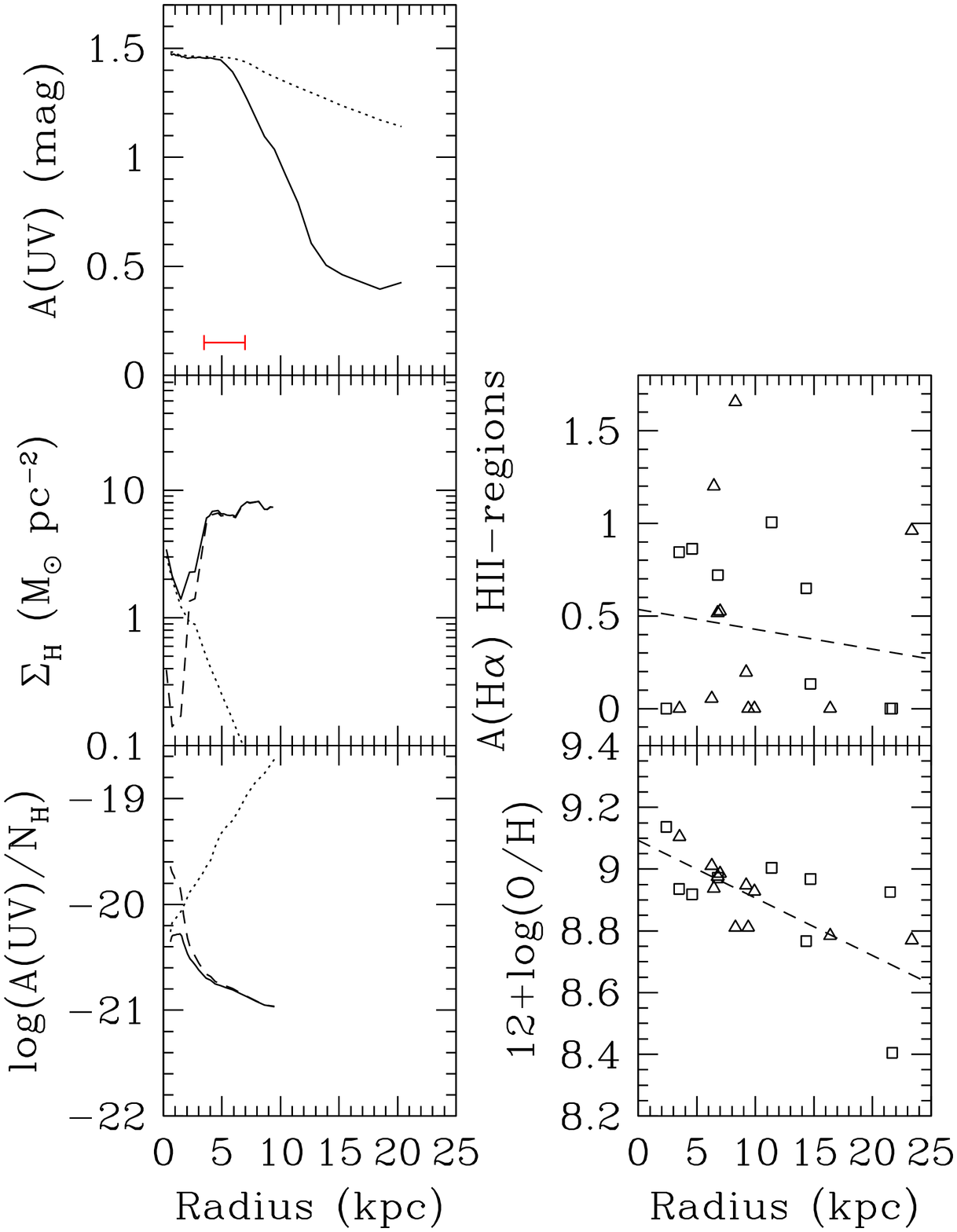} &
\end{tabular}
% \centering \includegraphics[width=0.5\textwidth]{individualM106.ps}
   \caption{Profiles in M106. 
Same caption as for Fig. \ref{figM33}.
} \label{figM106}%
\end{figure*}

The $A(H\alpha)$ extinction from the Balmer decrement for single HII regions
is shown in the 
\emph{middle row-middle column} panel of Figs.
\ref{figM33} to \ref{figM106}
(the number of points is smaller than the number of HII regions
given in Table 3 because we kept only those with corrections 
for the underlying Balmer absorption).
It shows large scatter at all radii.  A weak gradient
of decreasing extinction is nevertheless observed in all
the galaxies (although hardly in M81 and M106), as 
shown by the fit (dashed line).

For a few galaxies, 
independent measures of the dust extinction in HII regions are
available:

\citet{scoville01} used high resolution \Ha{} and \Pa{}
(Paschen-$\alpha$) images
of M51 to study the extinction in HII regions.
For illustration, we add their results to the \emph{top row-middle column} 
panel of Fig. \ref{figM51}. 
Each triangle corresponds
to an individual HII region, in which 
the extinction has been computed from the \Pa{}  to 
\Ha{} ratio.
On average, they find a larger extinction than  
the one derived from the Balmer decrement
at the same radius. 
This is partially due to a 
resolution effect, as \citet{scoville01} estimate
the extinction in very well defined regions in their \Pa{} and \Ha{} images.
The solid line shows the extinction computed from the \Pa{} to \Ha{}
ratio after the fluxes have been averaged in 10 arcsec radial bins
(only using the pixels where the signal to noise is larger than 3 and 5
for \Ha{} and \Pa{}). There is no sign of a gradient, but the 
data cover a limited radial range ($R<$ 4 kpc).

For M101, we present in Fig. \ref{figM101} as a shaded area
a sketch of the extinction variation with radius
found by \citet{scowen92} in 625 HII regions identified in
\Ha{} and \Hb{} narrow-band images. The scatter in their data is very
large (see their Fig. 3), but the trend is clearly detected.
Inside $\sim$ 20 kpc, they find a steeper gradient than 
the one obtained with the limited number of spectroscopic studies 
we consider, while
it is quite flat in the outer galaxy (where it is however based on only 
a few HII regions).

%%%%%%%%%%%%%%%%%%%%%%%%%%INDIVIDUAL FIGURES: BEGIN

%%%%%%%%%%%%%%%%%%END INDIVIDUAL FIGURES.

\subsection{Abundances in HII regions}

\begin{table}
\caption{References for complementary data, Position Angle ($PA$) 
and inclination ($incl$) in
degrees used for determining the profiles} 
\label{tabledatac}
\begin{tabular}{ l l l l r r  }
\hline
  & HI & CO & HII regions      & $PA$ & $Incl$ \\
M33  & 1 & 7 & 10,11,12(21)    &  22 & 55 \\
M51  & 2 & 8 & 10,13(10)       &  17 & 20 \\
M81  & 3 & 9 & 14,15,16(28)   & -28 & 59\\
M100 & 4 & 8 & 10,17(14)       & 153 & 27\\
M101 & 5 & 5 & 10,18,19,20(38) &  42 & 21\\
M106 & 6 & 8 & 14,21(20)         & -30 & 63\\
\hline
\end{tabular}

{\footnotesize
\emph{References for the HI.}
[1]: \citet{deul87}.
[2]: \citet{rand92}.
[3]: \citet{rots75}.
[4]: \citet{warmels86}.
[5]: \citet{wong02}.
[6]: \citet{wevers86}.
\emph{References for the CO.}
[7]: \citet{corbelli03}
[8]: \citet{young95}
[9]: \citet{sage93}
[5]: \citet{wong02}.
\emph{Spectral information for HII regions (for abundances and extinctions).}
The number between parentheses indicates the number of HII regions.
[10]: \citet{mccall85}.
[11]: \citet{kwitter81}.
[12]: \citet{vilchez88}.
[13]: \citet{diaz91}
[14]: \citet{oey93}.
[15]: \citet{garnett87}.
[16]: \citet{stauffer84}.
[17]: \citet{shields91}.
[18]: \citet{smith75}.
[19]: \citet{rayo82}.
[20]: \citet{vanzee98}.
[21]: \citet{zaritsky94}
}
\end{table}

\label{secabgrad}

The abundances are computed for the individual HII regions of 
each galaxy. The references for the HII regions data are 
given in Table \ref{tabledatac}. As most ot them were already collected by 
\citet{zaritsky94}
(and used to compute gradients),
the abundances of oxygen 
(12+log(O/H) are computed from the R23 
indicator ([OII]3726,3729+[OIII]4959,5007)/H$\beta$, using
their calibration.

For each galaxy, these abundances exhibit a clear 
gradient, shown as the dashed line in 
the \emph{bottom row-middle column} panel of Figs. \ref{figM33} to
\ref{figM106}.
This fit will be used hereafter to determine the abundance at
a given radius in each galaxy.
We implicitly make the assumption that the oxygen abundance measured in HII regions
is the same as in all the interstellar medium at the same distance from 
the centre of the galaxy.

In the case
of M101, we also show the oxygen abundance gradients
deduced by \citet{scowen92} from their analysis of 
625 HII regions (dotted lines). The different
curves correspond to different surface brightness thresholds that they
apply to select the HII regions. The advantage of the gradients of 
\citet{scowen92} is the large number of HII regions involved, 
although computed with limited spectral information. Contrary to other
studies, they propose a two slope gradient, but the abundances are nevertheless
close to those derived from the spectroscopy of the
limited number of HII regions of Table \ref{tabledatac}.
 
With high signal-to-noise spectra of 20 HII
regions, \citet{kennicutt03} recently derived the electron
temperature for those HII regions and determined robust abundances
which are systematically lower (by 0.2-0.5 dex)
than the abundances obtained with strong lines and empirical calibrators.
Figure \ref{figM101} shows that the \citet{kennicutt03} gradient is
also slightly steeper that from the strong-lines.

Since we do not have other data for the rest of our galaxies, we use
the abundance gradient as derived from the R23 calibration 
(including M101, for the sake of homogeneity).
If R23 systematically produces abundances that are too large,
our results should be corrected accordingly, perhaps by reducing our abundances
by a few tenths of dex.

The uncertainty in the abundances 
is usually on the order of $\sim$ 0.15 dex. This is much smaller than
the differences in the metallicity measured at various radii, so that
the value of the gradient is relatively well defined.

\subsection{Gas profiles}

The molecular profiles are computed from
the $I(CO)$ profiles; references are given in Table \ref{tabledatac}. 
When only a few points are available, an exponential profile
is fitted and used afterwards.
We use a conversion factor $X$ from $I(CO)$ 
to H$_2$ dependent on the metallicity, as suggested 
by \citet{boselli02}:
\begin{equation}
log X= -1.01 (12+log(O/H))+29.28.
\label{eqX}
\end{equation}
This correlation was found for integrated values over whole 
galaxies and is here assumed to hold for radial profiles as well. 
Note that \citet{nakai95} found similar results in M51
in their analysis of the radial variation of the dust-to-gas ratio.
The value of $12+log(O/H)$ used in Eq. \ref{eqX} is the one
given by the radial fit of the abundance gradient 
(see Sect. \ref{secabgrad}). 

References for the atomic gas profiles are given in Table \ref{tabledatac}.
Although these profiles 
are given at various resolutions, 
the results of our analysis should not be affected since
all our profiles are smoothed to the IRAS 100 $\mu$m resolution
in order to compare them with the derived UV attenuation.
For each galaxy, 
the gas profiles are shown in the \emph{middle row-left column} panels of Figs. 
\ref{figM33} to \ref{figM106}.

\subsection{FIR and UV profiles}

   \begin{figure}
   \centering
   \includegraphics[width=0.499\textwidth]{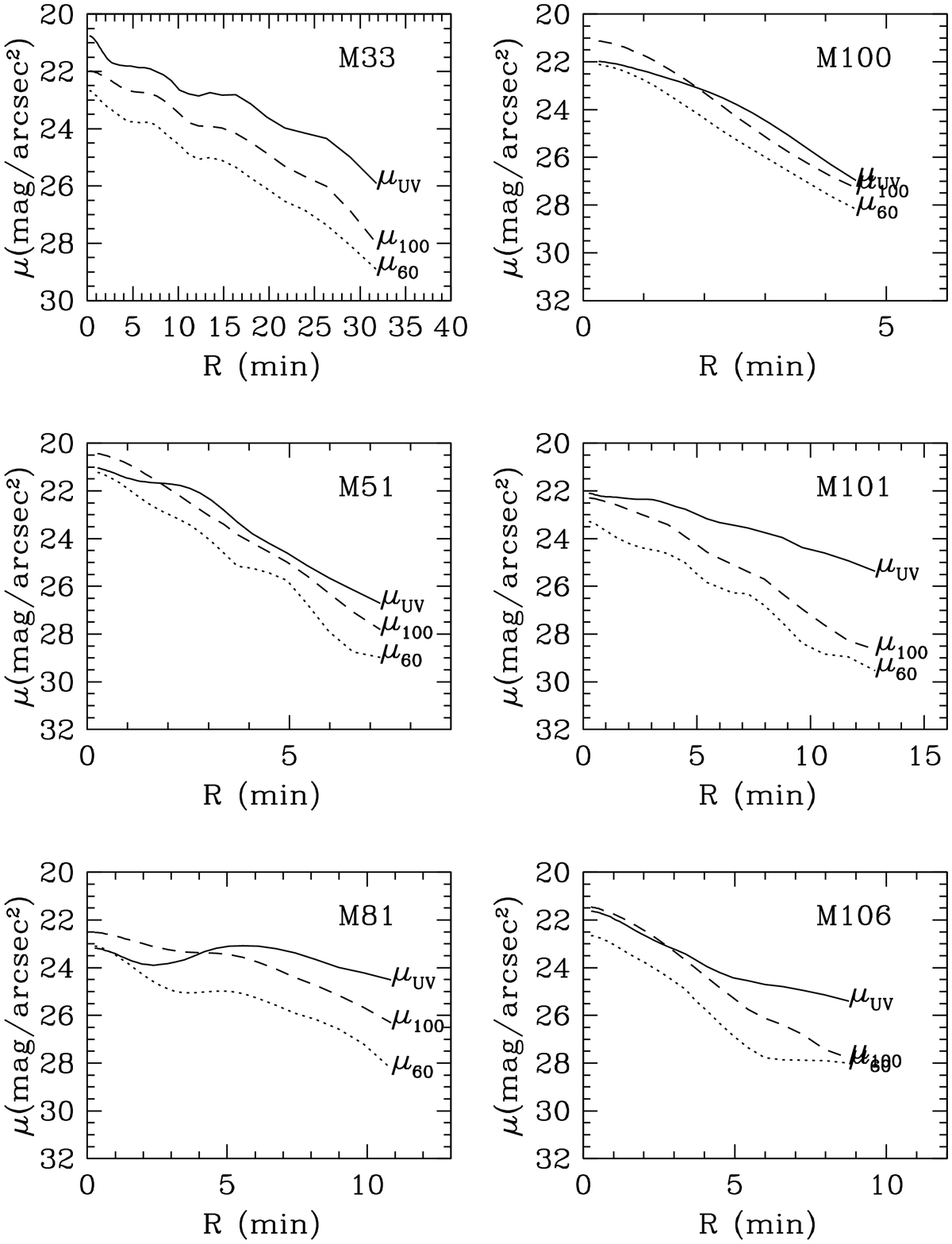}
      \caption{
Profiles of UV and FIR surface brightness at the IRAS 100 $\mu$m
resolution ($\sim$ 100 arcsec).
For the infrared, $\mu$ is defined as -2.5 log ($F$) + 15, $F$ in 
Jy arcsec$^{-2}$. In the UV $\mu_{UV}$=-2.5 log($F$)-21.175,
$F$ in erg cm$^{-2}$ s$^{-1}$ A$^{-1}$ arcsec$^{-2}$.
}
         \label{figprofuvfir}
   \end{figure}
From the IRAS and FOCA images, we compute the surface brightness
profiles at 2000 \AA, 60 and 100 $\mu$m as follows.  After removing
the stars present in the UV image and subtracting the sky, the
resolution in all images is changed to that of the 100 $\mu$m
maps (100 '') by convolution with a Gaussian filter. The pixel size
of the UV image is changed to match IRAS (15
arcsecs per pixel).

The fluxes are azimuthally averaged using the task ELLIPSE within
IRAF\footnote{IRAF is distributed by the National Optical Astronomy
Observatories, which are operated by the Association of Universities
for Research in Astronomy, Inc., under cooperative agreement with the
National Science Foundation.}, keeping the centre, ellipticity and
position angle fixed for each galaxy. We adopt the same position
angle and inclination as used in the determination of the 
HI profiles, and given in Table \ref{tabledatac}.

The UV profiles are corrected for Galactic extinction as taken
from \citet{schlegel98}.

The resulting surface brightness profiles are shown in Fig. \ref{figprofuvfir}
(for the FIR, we adopt $\mu$= -2.5 log ($F$) + 15, $F$ in Jy arcsec$^{-2}$).
We truncate the profiles at a radius where the azimuthally averaged 
flux surface densities 
become lower than
one sigma of the sky.

Our profile of M33 is in agreement with the one published by
Buat et al. (1994) at the FOCA spatial resolution, but shows
less structure because our data have been smoothed.

The profile of M106 shows a flattening in the outer part of the UV
and 60 $\mu$m flux. The feature must be real since we are well above
the sky in the UV and it is located in the area of a ring, clearly 
visible in the image.
At 60 $\mu$m, the flux at the same radius is barely more than 
one $\sigma_{sky}$, and is dominated
by a few large peaks, not present in the 100 $\mu$m image. 
The few outer points of this profile are quite uncertain.

\section{Radial extinction profiles}

The extinction profile $A(UV)$ is determined from the
FIR/UV ratio by combining the UV and FIR radial profiles (determined
as described in Sect. 2.4), using the calibration of 
Buat et al. (1999), in Sect. 3.1.
By combining these results with a simple geometrical model, we give
in Sect. 3.2 a simple recipe to extend A(UV) to
optical and NIR wavelengths.

\subsection{Determination of the UV extinction profile}

To compute the UV extinction $A(UV)$, we adopt an updated version of 
the fit of \citet{buat99} given in \citet{iglesias03}:
\begin{eqnarray}
%A(UV)=0.466+log \left(\frac{F_{FIR}}{F_{UV}}\right)+0.433 \left( log\left(\frac{F_{FIR}}{F_{UV}}\right) \right)^2,
A(UV)=0.622 &+ & 1.140 \ log \left(\frac{F_{FIR}}{F_{UV}}\right) \nonumber \\
            &+ & 0.425 \left( log\left(\frac{F_{FIR}}{F_{UV}}\right) \right)^2,
\label{eqabsratio}
\end{eqnarray}
where $F_{FIR}$ (W m$^{-2}$ arcsec$^{-2}$) is obtained from a combination 
of the 60 and 100 $\mu$m fluxes, i.e.:
\begin{equation}
F_{FIR}=1.26 (2.58 \, 10^{12} f_{60} + 10^{12} f_{100}) 10^{-26} 
\end{equation}
where $f_{60}$ and $f_{100}$ are the IRAS surface brightnesses in Jy
arcsec$^{-2}$; and $F_{UV}$ is the UV flux:
\begin{equation}
F_{UV}=2000 \,  10^{-3} f_{2000} [{\rm W m^{-2} arcsec^{-2}}]
\end{equation}
where $f_{2000}$ is the UV surface brightness 
(erg cm$^{-2}$ s$^{-1}$ \AA$^{-1}$ arcsec$^{-2}$). 

The IRAS FIR does not include the total dust emission. However,
the calibration of Iglesias et al. (2003), using the ISO (Infrared
Space Observatory) results of Dale et al. (2001), takes into account
the average difference between the FIR emission and the total dust
emission in disk galaxies.  
This extrapolation is nevertheless based
on data short-ward of 100 $\mu$m and model SED curves and does
not take into account the radial variation of the dust temperature.
Alton et al. (1998), using ISO,
found that the cold dust is indeed more extended than the warm dust in their
sample of disc galaxies. We assume that the energetic balance is not
much affected by this radial change, and implicitly use an average
calibration (note that our 60/100 ratio does not exhibit a large
gradient).
%Moreover the cold dust has important consequences
%for the dust mass (see section 5.2).}

\label{secrobustness}
The FIR to UV ratio is believed to be a robust 
method of determining $A(UV)$ \citep[e.g.][]{witt00}.
Other calibrations of the determination of the UV
extinction from this ratio have been proposed by
\citet{panuzzo03} for face-on and edge-on 
orientations.
As our galaxies are quite far from being edge-on 
(inclinations from 20 to 63 $^{\circ}$),
we compared our results with the $A(UV)$ obtained for
the face-on calibration of \citet{panuzzo03}. The difference
in A(UV) using the two methods is
less than 10 \%
except for extinctions lower than a few tens of magnitude,
as in the few outer points of M101 
%among our data 
(for a radius larger than $\sim$ 20 kpc, 
see Fig. \ref{figM101}),
confirming that the determination of A(UV) is mostly 
independent on the adopted calibration.

An implicit assumption in using the FIR/UV ratio is that
we take into account all the UV photons which may have 
heated the dust where it radiates the energy (after having 
eventually propagated inside the galaxy).
While this is not the case for regions as small as HII regions, 
it is certainly true for the integrated flux of galaxies.
In the resolved galaxies, this method can still be applied provided
that we average the flux over large enough areas, as for the
case of our sample where the azimuthally averaged profiles are
determined within 100 arcsec wide annuli (between 
$\sim$ 0.5 and $\sim$ 8 kpc in our galaxies), the IRAS
100 $\mu$m resolution.

In the \emph{top-left} panels of Figs. \ref{figM33} to \ref{figM106},
the resulting profile of the UV extinction $A(UV)$
is shown as a solid line (the
dotted line indicates the extinction integrated within the radius $R$,
obtained by applying Eq.
\ref{eqabsratio} to the integrated fluxes within the ellipses of
semi-major axis $R$).

For all the galaxies, we observe a clear UV extinction gradient
over a radial range several times the resolution. 
Only for M51 is a secondary
peak observed, whose radius corresponds to its companion galaxy.
Indeed the companion suffers very strong extinction 
since it is not visible in
the UV image while being prominent in FIR. The UV and FIR surface 
brightness
measured in the images at the position of the companion give a local
extinction of $\sim$ 3 magnitudes.

   \begin{figure} \centering
   \includegraphics[angle=-90,width=0.499\textwidth]{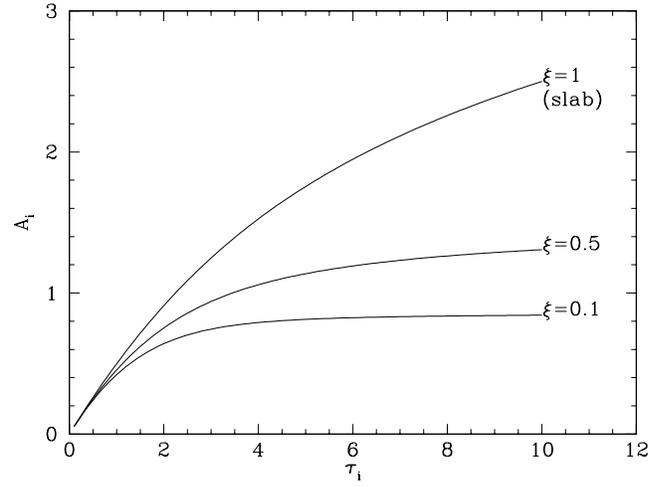}
   \caption{Relation between the absorption A$_i$ and the optical
   thickness $\tau_i$ for sandwich models of various dust to star
scale-height ratios $\xi$.}  \label{figabsotau} 
   \end{figure}

\subsection{A simple model to derive the optical thickness and the
extinction profile at various wavelengths}
\label{secmodext}

$A(UV)$ can be scaled to $A(\lambda)$ at any wavelength 
$\lambda$ once
an extinction law (Galactic, Magellanic) and a geometry
are assumed. To simplify the formalism and to be consistent
with previous studies \citep[e.g.][]{boselli03}, we assume a Galactic 
extinction law and a sandwich model
\label{secextmodel}
with a wavelength dependent dust-to-star scale-height ratio 
$\xi$. The extinction $A$ (in magnitude) then depends on
the optical thickness
($\tau$) via Eq. \ref{eqabsotau}:
\begin{equation}
\label{eqabsotau}
A_i=-2.5 log \left( \left(\frac{1-\xi}{2} \right) (1+e^{-\tau_i})
+ \left( \frac{\xi}{\tau_i} \right) (1-e^{-\tau_i}) \right).
\end{equation}
$\tau_i$, $A_i$, and $\xi$ are wavelength-dependent. $\tau_i$ and $A_i$ 
correspond to a given inclination, i.e. $\tau_i=\tau_0 sec(i)$.
The relation between $A_i$ and $\tau_i$ is shown in Fig. \ref{figabsotau}
for several values of $\xi$.

We numerically invert Eq.  \ref{eqabsotau} to derive the profile of 
$\tau_{UV}$ from the one of $A(UV)$. Then, we compute the
optical thickness at a given wavelength as : 
$\tau_{\lambda}=\tau_{UV} \  k(\lambda)/k(UV)$, where $k$
is a typical Galactic extinction curve ($R_V=A_V/E(B-V)$=3.1). 
The choice of this extinction curve is justified by the fact 
that the metallicity in our galaxies 
(see Table \ref{tablegen} and Fig. 
\ref{figdtgmetallin}, for instance) is larger than in the Magellanic Clouds.
We are implicitly assuming that the albedo does not depend
strongly on the wavelength and that the scattering is isotropic
(we also tried to include the effects of the albedo and phase function
as in \citet{calzetti94} with very similar results).

We can then compute the extinction at any wavelength
with the help of Eq. \ref{eqabsotau}. We show 
the result of this operation for the V and H band wavelengths
in Fig. \ref{figabsos}. 
We adopt a dependence of the dust-to-star scale-height ratio
on wavelength as suggested by \citet{boselli03}: 
\begin{equation}
\label{eqxivarying}
\xi(\lambda)=1.0867-5.501 \ 10^{-5} \lambda (\AA),
\end{equation}
This assumption is used to mimic the fact that the young 
stars (predominantly emitting at short wavelengths) lie in a thinner disk 
(similar to the dust, $\xi$=1 at 2000 \AA) 
than older stars (emitting at long wavelengths) which migrate
to larger heights with age. This is in agreement with some 
observations \citep{boselli94}, but not always confirmed
for edge-on nearby galaxies \citep[e.g.][]{xilouris99}.

Note that a sandwich model with a constant
dust-to-star scale-height ratio $\xi(\lambda)$=0.5 is not consistent
with the observations. It can be seen in
Fig. \ref{figabsotau} that for this value of $\xi$,
the absorption in magnitudes can reach only moderate values. 
Even for infinite optical thickness, 
we would obtain only $A(UV)$=1.5 which is lower than the absorption
measured in the centres of M51, M81 and M100.

The results of this model applied to our galaxies are shown in Fig.
\ref{figabsos} for the wavelengths of the V and H bands.

The variation of extinction with radius that 
we can detect is between $\sim$ 0.5 and $\sim$
1 mag over the whole radial range in the UV. 
The results of the model indicate that the
typical V band extinction gradient is between 0.2 and 0.5 mag,
and it is very small in the H band. Of course this is model-dependent
and could easily vary by a few tenths of magnitude in the V band
if we change $\xi$.  

Fig. \ref{figabsos} also shows the extinction profiles
in V and H when
a simple screen of Milky Way-like dust is adopted (marked ``S'' in the figure) 
or with the \citet{calzetti99} law (``C''). 
The differences between these models are negligible at
long wavelengths but can reach 0.3 magnitude in V (and eventually change
with the radius).

This shows that the geometry (and the choice of one geometrical and dust model
rather than another) can have an important impact on the determination
of the colour gradient of the underlying stellar population, even if
the observed colour gradient is unlikely to be mainly due to the dust
\citep{dejong96}.
The extinction dependence on the geometry is a well known effect,
studied in detail in Disney et al. (1989).

\subsection{Comparison with other studies of M51}

\citet{hill97} show a UV colour profile of M51 
($\mu_{151}$-$\mu_U$) varying by 1 mag or so between
$\sim$ 2 and 12 kpc (the gradient is even stronger if smaller
radii are included, but we cannot probe such small scales
with our resolution). Hill et al.
propose that the colour gradient is due
to an extinction gradient.  We indeed find (in a more direct way)
an extinction gradient in the UV in M51 (Fig. \ref{figM51}).

\begin{figure*}
   \centering
   \includegraphics[angle=-90,width=\textwidth]{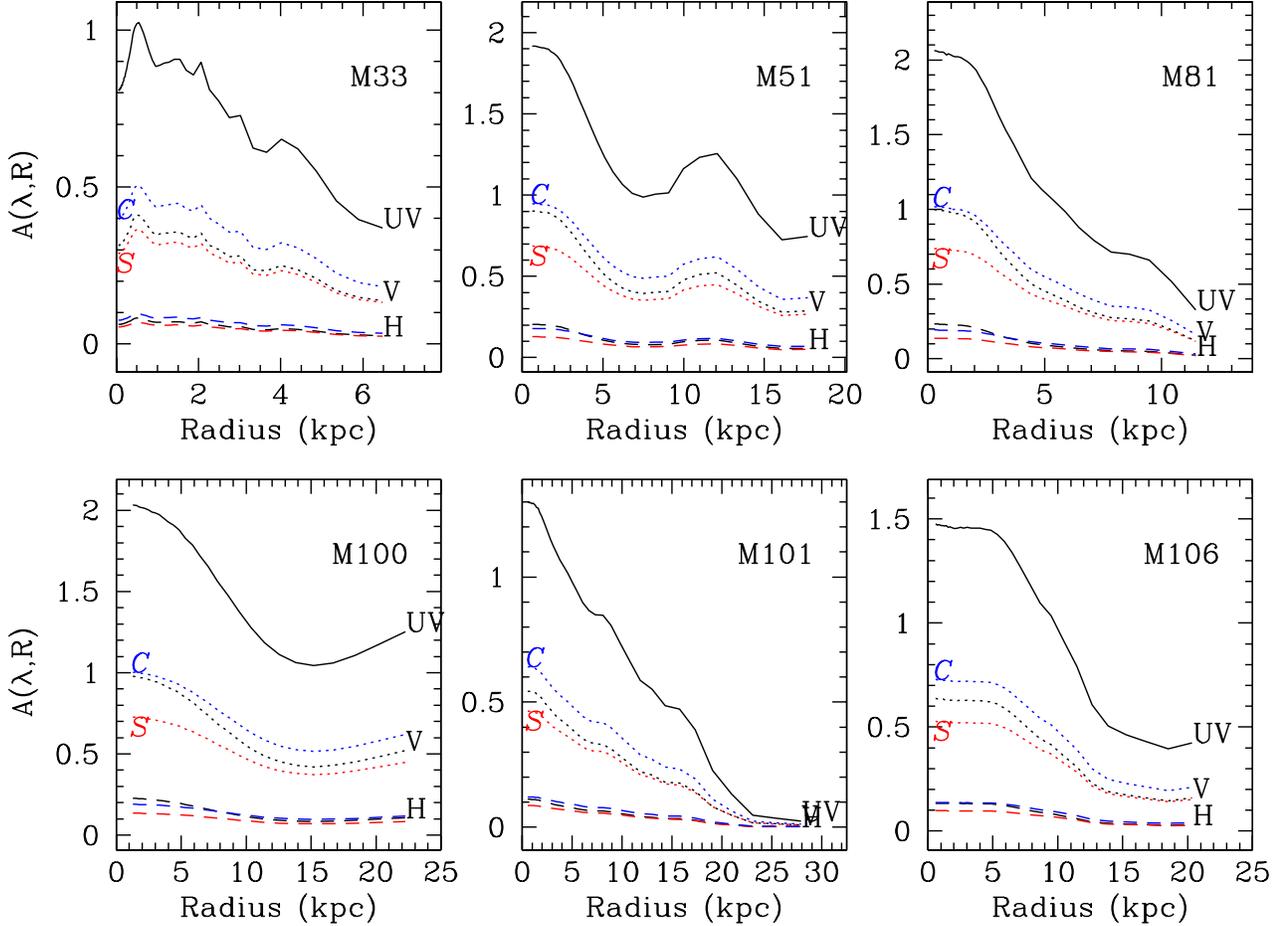}
      \caption{Extinction profiles in the UV, V and H.
$A(UV)$ is obtained from the FIR/UV ratio. The black line
shows the extinction profiles in V and H as derived for a sandwich model with 
dust to star scale-height ratio $\xi$ depending on the wavelength.
For each of these two bands, the others curve show the predicted extinction
for a dust screen with Milky-Way-type dust (S) and the 
\citet{calzetti99} law (C). 
} 
         \label{figabsos}
   \end{figure*}

\citet{vansevicius01} estimates the extinction
at 11 kpc from the centre of M51 using its companion as a tracer. He
finds A$_V$ = 0.7 $\pm$ 0.3 mag. The extinction we derive from our model
at this radius is a bit lower ($\sim$ 0.5 mag) but within the error bars (see
Fig. \ref{figabsos}). As another illustration of the dependence of our
results on the adopted geometry, changing $\xi$ to a constant value of
0.7 (independent of the wavelength) would increase the value of A$_V$
by only 0.1 magnitude (larger differences would occur where the
optical thickness is larger, i.e. in the inner few kpc of the galaxy).

\section{Extinction in the UV continuum and in HII lines: A(\Ha) vs A(UV)}

To check the consistency of our extinction determination with
average values available in the literature, we estimate the total
integrated extinction for both the UV and \Ha{} and
compare them to the results of \citet{buat02} for late-type normal 
spirals.

The integrated UV extinction within the radius where it is
computed can be seen as the dotted curve in the top-left panels
of Figs. \ref{figM33} to \ref{figM106}.  
It usually presents a plateau or a small decrease
at large radii; we then consider as ``integrated'' extinction
over the whole galaxy
its value at the largest radius.
%%%%%%%%%%%%%%%%%%%%%%%%%%%%% ??? XXX===XXX===
%For M33, we obtain the value of 0.54, which is consistent with
%the estimate ($\sim$ 0.4) of \citet{buat94}, using a method
%based on the HI content of the galaxy, and another one on the 
%total diffuse IR emission.
We choose as a typical H$\alpha$ extinction 
the average of A(\Ha) found in the HII regions
considered in this study.

\begin{figure} \centering
\includegraphics[width=0.499\textwidth]{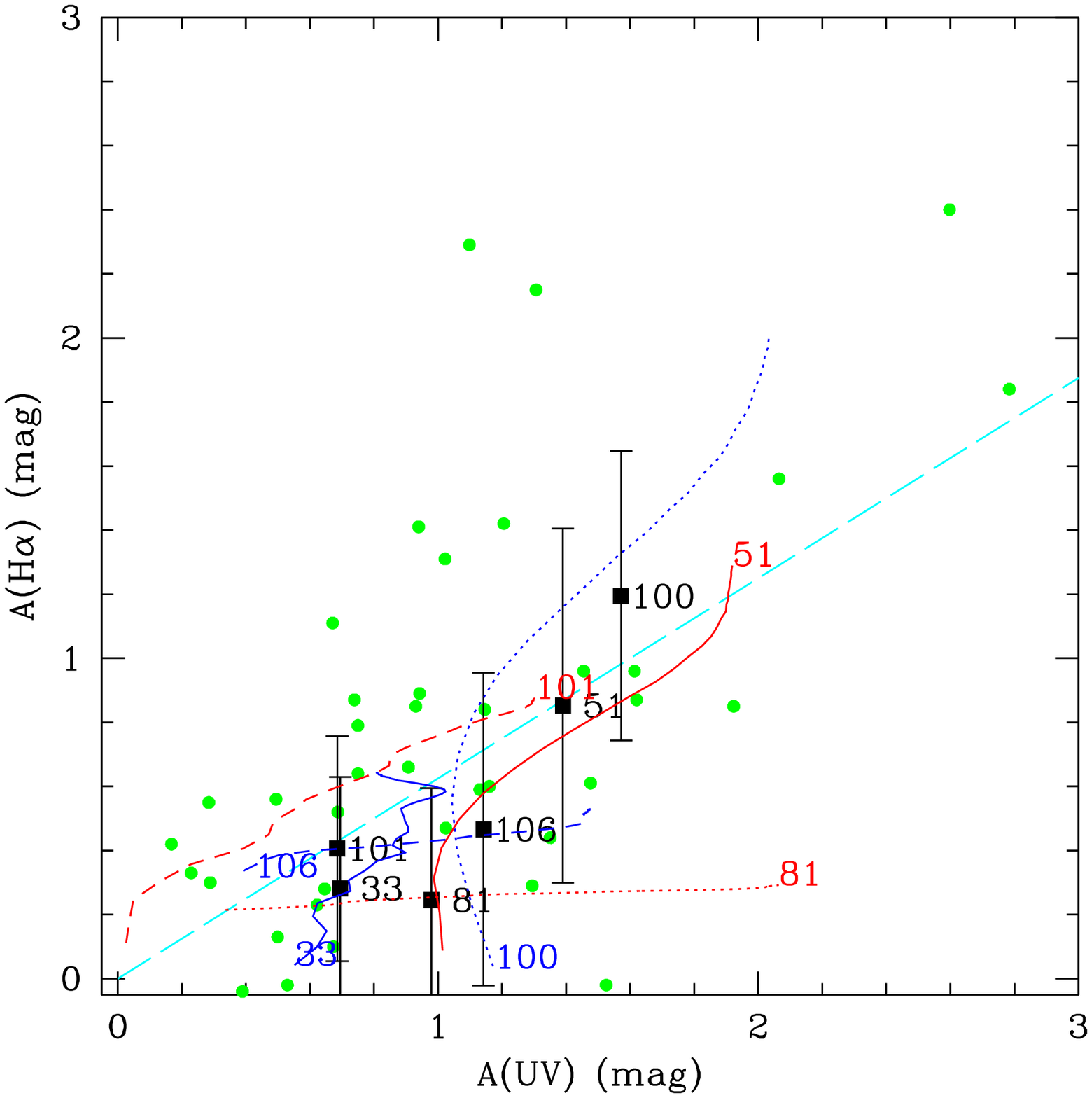} 
\caption{UV vs
   H$\alpha$ extinction.  The squares are the average values for
   each galaxy. The curves are local values at different radii
   within each galaxy. Filled circles are the 
   star-forming galaxies of \citet{buat02} \citep[adopting the $A(H\alpha)$ 
of][]{gavazzi03}, 
computing $A(UV)$ from the
   UV/FIR ratio, like for our galaxies, but using integrated fluxes.  
   The long dashed line indicates the relation $A(UV)=1.6 A(H\alpha)$,
   expected in the case of the attenuation law
   proposed by \citet{calzetti97} for starbursts.
}  
\label{figAuvAha}
   \end{figure}

The average H$\alpha$ extinction (and its standard deviation) 
and the UV integrated extinction are compared in
Fig. \ref{figAuvAha}.  
This figure also shows the local values (at different radii)
within each galaxy connected by
a curve. The integrated value is representative of an ``average''  
of the local values obtained at different radii.
The filled circles are the star-forming galaxies of \citet{buat02},
using the updated \Ha{} extinction of \citet{gavazzi03} 
(which introduces
small differences with respect to Buat et al.).

Our ``average'' values (squares) and most of the profiles (curves)
largely overlap the points
of \citet{buat02}, however,
we span
a limited range of extinction 
than they do
since we are limited to absorptions less than 2 mag, while Buat et al.
include objects with up to $\sim$ 4 magnitudes. Actually, this is the case of M100
for which Buat et al. find $A(H\alpha)>$ 3.7 mag.  However, they
notice that this high value may be due to the very high surface
brightness nucleus while the disk might be less extincted.  Our
H$\alpha$ extinction derived from HII regions in the disk is indeed
lower.

An important difference between our approach 
and that of \citet{buat02} is
that we defined an average $A(H\alpha)$ 
in HII regions, while
their integrated value is
dominated by high surface brightness regions and the nucleus. 
This might explain the differences between the two studies
and especially why our ``local'' study does not
reach the high values found by \citet{buat02} in ``integrated'' 
galaxies.

We can also see
in Fig. \ref{figAuvAha} that lower values of the integrated UV
extinction are found for
M33 and M101 (which are of type Scd) than for the other galaxies (type
Sbc and Sab). The average values for these two groups are in agreement
with (albeit slightly lower than) the average values given by
\citet{boselli03} for Sc-Scd (0.85 mag), and for Sa-Sbc galaxies (1.28 mag),
obtained for a much larger number of integrated galaxies.

The profile of M81 in $A(H\alpha)$ is rather flat, as seen
in Fig. \ref{figAuvAha}. It can be seen in Fig. \ref{figM81}
(\emph{middle row-middle column} panel) that this is due to a large
scatter of the extinction in a limited number of HII regions.
In general, the average values of $A(H\alpha)$ 
(squares in Fig. \ref{figAuvAha}) are more
robust than the local values as the latter are more sensitive
to the small number of HII regions.

Finally, we also show in Fig. \ref{figAuvAha} the
\citet{calzetti97} attenuation law for starbursts
\citep[$A(UV)=1.6 A(H\alpha)$, see][for the derivation]{buat02}
The relation is steeper for the integrated star-forming galaxies, 
as shown in \citet{buat02}, although consistent, within the large
uncertainties, with local values.

\section{Extinction, metals, and gas}

\subsection{Dependence of the extinction on the metallicity}

\citet{quillen01}, using extinction derived from the Paschen-$\alpha$ (\Pa) to H$\alpha$
ratio in the centres of a few galaxies and along the profiles of M51 and
M101 \citep[with original data from the thesis of ][]{scowenthesis}, 
found that the extinction correlates with the metallicity and
with the gas surface density (their Fig. 4). 
\citet{heckman98} have shown a dependence of the UV spectral slope $\beta$
on O/H in their starbursts galaxies
($\beta$ is defined as $F(\lambda) \propto \lambda^{\beta}$ for 
$1250<\lambda<2600$ \AA{} by \citet{calzetti94} and is equal to $\sim$ -2.1 
for a dust-free starburst).
Several studies, theoretical
and observational, have tried to relate the amount of extinction to metallicity 
and gas amount
\citep[e.g.;][]{issa90,lisenfeld98,dwek98,hirashita01}. 

From eq. \ref{eqabsotau} (adopting the varying dust-to-star
scale-height $\xi$, as given in Eq. \ref{eqxivarying}) we derive for each
profile of $A(UV)$ the corresponding profile of $\tau_{UV,i}$, which is
corrected for inclination, to obtain $\tau_{UV,0}$.

\begin{figure}
   \centering
   \includegraphics[width=0.499\textwidth]{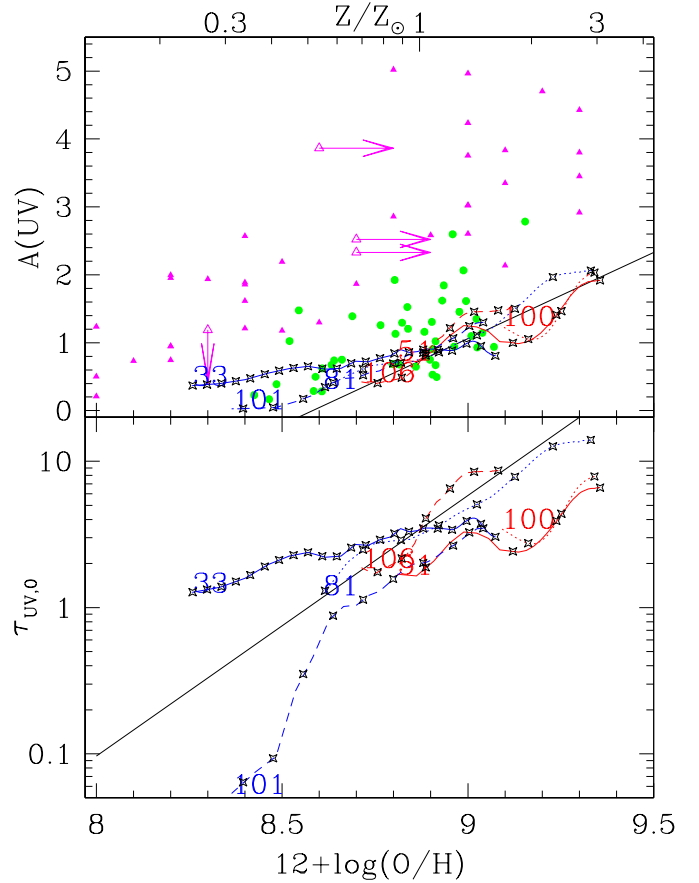}
      \caption{UV Extinction (top) and optical thickness (bottom)
as a function of 12+log(O/H). Each
galaxy is represented by a curve. The symbols correspond to radii
separated from each other in units of the spatial resolution. The 
median of a fit performed on each galaxy is shown as black line.
In the top panel, we also show the extinction derived for the
integrated starbursts of \citet{heckman98} as triangles
and the star forming galaxies of \citet{buat02} as circles.
              } 
         \label{figtaumetal}
   \end{figure}

In Fig. \ref{figtaumetal} (bottom), we superpose the relation between
the optical thickness $\tau_{UV,0}$ and the abundance for all the
galaxies. In the top panel, we directly plot the extinction $A(UV)$ in
magnitudes (uncorrected for inclination) as a function of the oxygen
abundance.  The advantage of using $A(UV)$ as derived from the FIR to
UV flux ratio with respect to other dust extinction indicators is that
it does not depend on the extinction model.
%%% VERO
A quite good correlation between the extinction and the metallicity is observed, 
suggesting that the extinction does depend on the metallicity. 
This dependence is confirmed by the UV extinctions of the
integrated star forming galaxies
of \citet{buat02}, shown as circles, adopting the metallicities
of \citet[][and in preparation]{gavazzi03}, although with a larger
scatter. This correlation between local extinctions and metallicities
amongst our different galaxies
is a striking result as the oxygen
abundance and $A(UV)$ are obtained 
independently of
each other.

Actually, a dependence of the UV extinction on the metallicity
was already suggested by \citet{heckman98} for their starburst
galaxies. We show their data as triangles in the same figure.
Following \citet{buat02}, the $A(UV)$ extinction at 2000 \AA{}
in starbursts is computed as
\begin{equation}
A(UV)=0.9 \times 2.5 \, \, log \left( \frac{1}{0.9} \, \frac{F_{FIR}}{F_{UV,1600 \AA}}+1 \right).
\end{equation}
The extinction in starbursts is much larger for the same metallicities
and shows a larger scatter than what we find within 
the disks of our spirals, as
already remarked by \citet{buat02}.

The black lines are the median of the fit
performed for each of our galaxies. The numbers given in parentheses in eq. 8
indicate 
the standard deviation in the parameters found for each galaxy.
These numbers reflect the large differences between galaxies, nevertheless
the global trend is well established.
 \begin{eqnarray}
log(\tau_{UV,0}) & = & -15.25 (\pm 6.27) + 1.78 (\pm 0.70) \times Z \nonumber \\
or: \label{eqfitz}  \\
A(UV)& = &  -21.80 (\pm 11.70) + 2.54 (\pm 1.28) \times Z \nonumber 
 \end{eqnarray}
where $Z=12+log(O/H)$. The standard deviations of 
$\tau$ and $A(UV)$ around the median-fit
are $\sigma$=0.37 and 0.43 respectively.
% SIGMA = 0.3409270346

The large scatter produces an important uncertainty in the zero-point
of this relation, even though the dependence on the metallicity is well defined.
These relations could be used to predict $\tau$ (or $A(UV)$) as a function
of the radius for each galaxy with an available abundance gradient.
In the case where no abundance gradient is available, we can use
the fact that the abundance gradient is constant in units
of the disc scale-length 
\citep[][and references within]{prantzos2000, henry99}. 
The oxygen gradient is typically -0.2 dex / $R_B$ ($R_B$
is the disk scale-length in the blue band).
The zero point of the relation can be determined from
the magnitude-metallicity relationship. 
\citet{garnett02} has shown that the oxygen abundance at the
effective radius in spiral galaxies satisfies:
\begin{equation}
\label{eqmassmetal}
[log(O/H)]_{Reff}=-0.16 M_B - 6.4
\end{equation}
where $M_B$ is the blue absolute magnitude. For an exponential disk,
$R_{eff}=1.865 R_B$. Combining this result with Eq. \ref{eqfitz},
we obtain a second expression of the optical thickness 
with no explicit dependence on the abundance gradient:
 \begin{eqnarray}
log(\tau_{UV,0}) & = & -0.28 (\pm 0.11) M_B -0.36 (\pm 0.14)  \frac{R}{R_B}  \nonumber \\
                 & - & 4.62 (\pm 10.45) \nonumber \\
or: \label{eqfitz2}  \\
            A(UV)& = & 0.40 (\pm 0.20) M_B  -0.51 (\pm 0.25) \frac{R}{R_B}  \nonumber \\
                 & - & 6.60 (\pm 19.38)  \nonumber  
 \end{eqnarray}

These relations can be used for any galaxy with an absolute magnitude
$M_B$ and scale-length $R_B$. The uncertainties given in Eq.
\ref{eqfitz2} are obtained only by
propagating the uncertainties of Eq. \ref{eqfitz}. 

To check the validity of Eq. 8 for computing an extinction
profile from the abundance gradient, in the top-panel of figure 12,
we show as a function of radius
the difference between the extinction
deduced from the FIR/UV ratio ($A(UV)_{FIR/UV}$) and the one deduced
from the oxygen abundance at this radius ($A(UV)_{O/H}$).  The
bottom panel shows the ratio of the two optical thicknesses.
   \begin{figure} 
	\centering
	\includegraphics[width=0.5\textwidth]{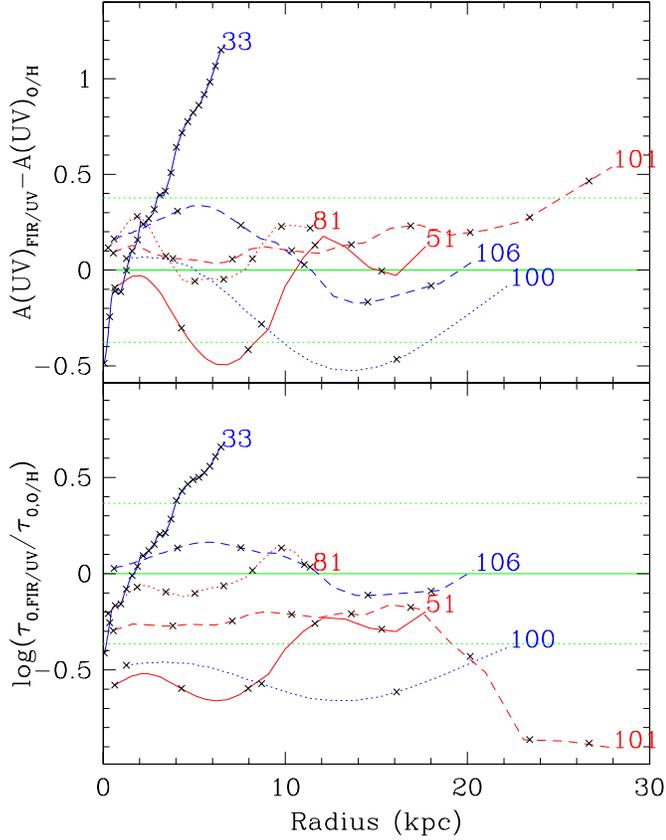} 
\caption{\label{figresidu} Top: Radial profiles of the difference between $A(UV)$ deduced 
from the FIR/UV ratio and $A(UV)$ obtained by 
applying Eq. 8 to the oxygen abundance.
Bottom: radial profile  of the  
ratio of the optical thickness deduced from the FIR/UV ratio
to the one obtained by applying Eq. 8 to the oxygen abundance.
The horizontal lines show the
average value and $\pm$ one $\sigma$. }
\end{figure}

The optical thickness ``predicted'' by applying Eq. 8 to the
oxygen abundance, $\tau_{0,O/H}$, is within a factor two of the
``observed'' one $(\tau_{0,FIR/UV})$ for most of the points.
Large departures are found only for the case of M33, which is the
galaxy with the smallest metallicity in our sample.

Similarly, the $A(UV)_{O/H}$ deduced from the metallicity is
frequently found at $\sim$ 0.3 mag from $A(UV)_{FIR/UV}$ and the
largest deviation is for M33 (up to more than one magnitude in its
outer parts). This may indicate that our relations are valid only in
the high-metallicity domain.

The points situated at radii larger than $\sim$ 20 kpc also show 
large differences between the predictions and the observations.
However, these data only concern the outer part of one galaxy
(M101) where $A(UV)$ is very small anyway (and where the 
calibration of our method is not as secure: see Sect.
\ref{secrobustness}). In any case, deviations from the average 
relation in the external parts are expected since the
uncertainties on the profiles (especially in the FIR) are larger.

%%%%%%%%%%%%%%%%%%%%%%%%%%%%%%%%%%%%%%%%%%%% SECTION GAS RATIO.
\subsection{Dust-to-Gas ratio}

The extinction-to-gas ratios (\emph{bottom-left} panel of Figs. \ref{figM33} to \ref{figM106}) 
are computed from the gas profiles (\emph{middle row-left column}), 
after smoothing the resolution to the IRAS 100 $\mu$m maps. This ratio 
shows various values and trends for each galaxy.

Based on extinction measured in HII regions at various radii,
\citet{issa90} suggested that the dust-to-gas ratio
depends on the metallicity.
This is quite natural since the dust is formed from
metals and the mass of metals is equal to the product 
of the metallicity and the gas mass.
This trend is naturally  obtained in the framework
of models computing consistently the evolution
of metals and dust, despite the large uncertainties 
in the yields of both \citep[e.g.][]{inoue03,dwek98}.

\citet{guiderdoni87} used the the solar neighborhood and the Magellanic Clouds
to propose the relation $\tau_{\lambda}/N_H \propto (Z/Z_{\odot})^s$
with s $\sim$ 1.35 (1.6) for wavelengths shorter (longer) 
than 2000 \AA, ($N_H$ represents the total hydrogen column density).

\begin{figure}
   \centering
  \includegraphics[width=0.5\textwidth]{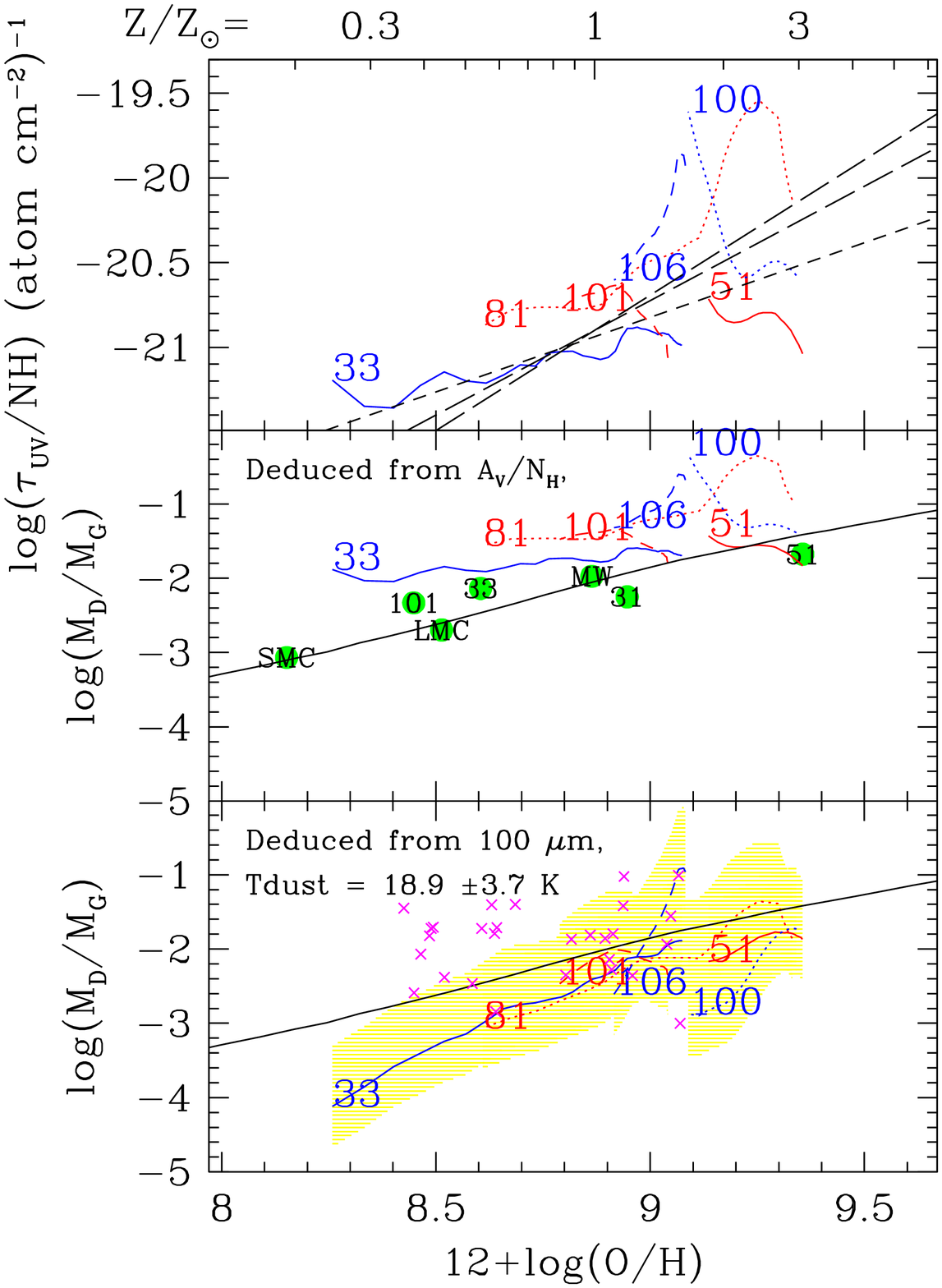} 
\caption{\emph{Top}: Optical thickness to hydrogen column density ratio.
The short-dashed line is a fit to our data of slope 0.88. The
long-dashed lines show the metallicity dependence of the dust to gas
ratio of \citet{guiderdoni87}.
\emph{middle}: dust to gas mass ratio (computed from the $A_V/N_H$ ratio as
explained in Sect. \ref{secmassratio}) as a function of metallicity. 
The circles mark the integrated galaxies of \citet{issa90}.
The solid curve is the best-fit model of \citet{hirashita01}. 
\emph{bottom:} dust to gas mass ratio (computed from the 100 $\mu$m flux
density). 
The crosses indicate the integrated galaxies of \citet{tuffs02}
and \citet{contursi01}.
The solid curve is the best-fit model of \citet{hirashita01}.
    \label{figdtgmetallin}
}   
\end{figure}

As can be seen in Fig. \ref{figdtgmetallin} (top), we obtain
a similar trend, although with a large dispersion,
especially for the metal-rich part.
% in the sense that,
%on average,
%larger optical-thickness-to-gas
%ratios %(proportional the dust to gas ratio)
%are found for larger metallicities. No large values
%of this ratio are found in the less metal-rich regions.
%In metal-rich regions, however, the dispersion is quite large.
A formal fit (short dashed line in Fig. \ref{figdtgmetallin})  
to our data gives a slope of 0.88. 
This is lower than the
values proposed by \citet{guiderdoni87}, which are illustrated by 
the two long-dashed lines in Fig. \ref{figdtgmetallin}.
Note however that we apply different methods 
(FIR/UV ratio, metallicity dependent CO to H$_2$ conversion factor...)
and probe
different ranges of metallicities as our spirals are more metal-rich
than the Magellanic Clouds.

\label{secmassratio}
Following \citet{hirashita01}, we can convert our optical-thickness-to-hydrogen 
ratio into a dust-to-gas mass ratio. This is done by
using the profiles of $A_V$ (Fig. \ref{figabsos}, computed with the
model of Sect. \ref{secextmodel}) and assuming that
$M_D$/$M_G$ (the dust-to-gas mass ratio) is equal to 
6 $\times$ 10$^{-3}$ when $N_H/E(B-V)$=5.9 10$^{21}$ 
\citep[][and references within]{hirashita01}.
The results are shown in the middle panel of Fig. 
\ref{figdtgmetallin}. 
In this figure, we compare our results with the 
dust-to-gas mass ratio found by \citet{issa90} for 6 galaxies.
%with independent methods. 
They were deduced from the 
$A_V/N_H$ of HII regions (or galaxy counts and interstellar 
absorption in the Milky Way), estimated at a radius of 0.7 $R_{dV}$
\citep[$R_{dV}$ is from][]{vaucouleurs76}.

Three of the galaxies of \citet{issa90} are in common with our
study (M33,M51,M101), presenting some differences with our results
that must come from the fact that we derive $A_V$ from the
FIR/UV ratio and use the gas profile, while they use measurements done
in HII regions. In M101, our metallicity is higher, but our gradient
is based on several, more recent studies.

A more direct way to estimate the dust-to-gas mass ratio is 
to deduce the dust mass from the FIR flux. We compute the dust
mass following \citet{devereux90} with the same numerical coefficient
as in \citet{boselli02}:
\begin{equation}
M_D= 1.27 F_{100} D^2 (exp(144/T_D)-1) M_{\odot}.
\end{equation}
The mass is very sensitive to the dust 
temperature $T_D$ which is still poorly constrained
by the 100 $\mu$m IRAS data. In the bottom-panel of Fig. 
\ref{figdtgmetallin}
we show as crosses the dust-to-gas mass ratio 
obtained from the ISO measurements of 
integrated flux densities of galaxies by 
\citet{tuffs02} and \citet{contursi01}, 
where the metallicity is from \citet{gavazzi03}.
We adopt the average temperature ($T_D$=18.9 K) of the sample
of \citet{tuffs02} (for which the dust temperatures are given in
Popescu et al., 2002) to compute
the dust-to-gas mass ratio of our
galaxies at different radii. 
This temperature is consistent with the 200 $\mu$m observations of
Alton et al. (1998).
The result is shown in the same
figure as a function of the metallicity. The shaded area
indicates a dispersion of $\pm$ 3.7 K.

The dust-to-gas ratio is lower than the one deduced from
the $A_V/N_H$ and shows a clearer trend with metallicity,
still with a large dispersion.

If a dust temperature gradient is present (with higher $T_D$ in the
center of galaxies than in their outskirts), as expected from the
observed gradient in metallicities and star formation rate (and
suggested by the ISO observations of e.g. Alton et al., 1998), it would
flatten the relation between the dust-to-gas mass ratio and the
metallicity, making it more similar to the one derived from the
$A_V/N_H$ ratio (see bottom panel).

The best-fit model of \citet{hirashita01} is 
indicated in the figure by the solid curve.
It reproduces the global trend of increasing dust-to-gas ratio
from dwarfs \citep{lisenfeld98} to spirals \citep{issa90}.

When a dust-to-gas ratio is 
determined as in \citet{issa90}, the trend for our galaxies
is weaker than in this model, 
and a great deal of diversity is observed
amongst the galaxies. The values obtained from the 100 $\mu$m
surface brightness are in slightly better agreement with the 
model. The uncertainties due to the temperature (shaded
area) and its gradient (see discussion above) are however 
large.

\section{Conclusion}

We have combined 2000 \AA{} UV images obtained with FOCA, and FIR IRAS
images at 60 and 100 $\mu$m to compute the FIR and UV profiles of six
nearby late-type galaxies.  We used the FIR/UV ratio to trace the
radial variation of the UV extinction in each galaxy.  We detect a
monotonic gradient of decreasing extinction with the radius in all of
them, except in M51 where the companion produces a second peak in the
averaged profile.

These extinction profiles were compared to the extinctions derived
from hydrogen lines (mainly the Balmer decrement) in HII regions and
we studied the relation between the extinction, the gas surface
density, and the metallicity.

The most significant result of this analysis is a clear correlation
between the UV extinction (in magnitudes or in optical thickness) and
the metallicity deduced independently from the FIR/UV profile and
the abundance gradient respectively.  This correlation is also found
in the integrated star forming galaxies of \citet{buat02} (a similar
relationship was found in starbursts \citep{heckman98}, but with
larger scatter and higher extinction for a given metallicity).

A fit to our azimuthally averaged data provides a simple relationship
between the extinction in the UV and the metallicity in galaxies for
which an abundance gradient is available (Eq. \ref{eqfitz}).
Coupling this result with the mass-metallicity relationship, we
derived an empirical formula linking the extinction profile to the
blue absolute magnitude and disc scale-length of galaxies (Eq.
\ref{eqfitz2}). 
Once the UV extinction profile is
determined by one of the previous methods, the extinction profile at
any wavelength can easily be derived through a simple model as the one
described in Sect. \ref{secmodext}.

Our results were obtained with a very limited sample of
galaxies. In the future, we will however be able to extend this
work with the new images of GALEX in the UV and SIRTF in the
far-infrared for a larger number of galaxies. 
GALEX data are being used to study the extinction radial profile in
individual spirals: in M83 (Boissier et al., 2004) and in a comparative
study with ISO FIR maps for M101 (Popescu et al., 2004).

\section*{Acknowledgement}

This research has made use of the NASA/IPAC Extragalactic Database
(NED) which is operated by the Jet Propulsion Laboratory, California
Institute of Technology, under contract with the National Aeronautics
and Space Administration. S.B. thanks the CNES for its financial support
and the Carnegie staff for welcoming him in Pasadena, and particularly
B. Madore and A. Gil de Paz for their comments on this work; 
M. Polletta for providing data and useful discussions.
A.B. thanks B. Madore for inviting him to the Carnegie Observatories
in Pasadena. We also thank G. Gavazzi and A. Zaccardo for providing
the metallicity of the Virgo galaxies.

FOCA has been funded by the Centre National d'Etudes Spatiales
and Fonds National de la Recherche Scientifique.

\def\apj{ApJ}
\def\aap{A\&A}
\def\aaps{A\&AS}
\def\apjs{ApJS}
\def\mnras{MNRAS}
\def\aj{Astronomical Journal}
\def\apss{Ap\&SS}

%\newpage
%FOR THE AUTHORS' INTEREST.
%\tableofcontents

\end{document}